\newif\ifsingle
\twocolumn

\newif\ifFullVersion
\FullVersiontrue 

\ifsingle
\documentclass[12pt,draftclsnofoot, onecolumn]{IEEEtran}		
\else		
\documentclass[10pt,final, twocolumn]{IEEEtran}
\fi

\usepackage{times}
\usepackage{amsmath,dsfont}
\usepackage{amssymb,amsthm}
\usepackage{epsfig,verbatim}
\usepackage{subfigure,stfloats}
\usepackage{setspace}
\usepackage{color}
\usepackage{cite}
\usepackage{epstopdf}
\usepackage{graphics}
\usepackage{accents}
\usepackage{acronym}
\usepackage[bookmarks,colorlinks]{hyperref}
\usepackage{booktabs}
\usepackage{mathtools}
\usepackage{algorithm}
\usepackage{algorithmic}
\usepackage{cuted}

\usepackage{enumitem}
\usepackage{bbm}

\newtheorem{proposition}{Proposition}

\ifsingle

\else

\fi 

\acrodef{csi}[CSI]{channel state information}
\acrodef{dma}[DMA]{dynamic metasurface antenna}
\acrodef{sinr}[SINR]{signal-to-interference-and-noise ratio}
\acrodef{bs}[BS]{base station} 
\acrodef{em}[EM]{electromagnetic} 
\acrodef{mimo}[MIMO]{multiple-input multiple-output}
\acrodef{ris}[RIS]{reconfigurable intelligent surface}
\acrodef{awgn}[AWGN]{additive white Gaussian noise} 
\acrodef{ula}[ULA]{uniform linear array}
\acrodef{upa}[UPA]{uniform planar array}
\acrodef{dfrc}[DFRC]{dual-function radar-communication}
\acrodef{hris}[HRIS]{hybrid \ac{ris}}
\acrodef{fgs}[FGS]{fast grid search}
\acrodef{agd}[AGD]{auto gradient descent}
\acrodef{rf}[RF]{radio frequency}
\acrodef{fov}[FOV]{field of view}
\acrodef{ga}[GA]{genetic algorithm}
\acrodef{sdp}[SDP]{semidefinite programming}
\acrodef{sdr}[SDR]{semidefinite relaxation}
\acrodef{ad}[AD]{auto-differentiation}
\acrodef{ao}[AO]{alternating optimization}

\IEEEoverridecommandlockouts



\title{Hybrid RIS-Assisted MIMO Dual-Function Radar-Communication System
}
\author{  
	\IEEEauthorblockN{Zhuoyang Liu,~\IEEEmembership{Graduate Student Member,~IEEE}, Haiyang Zhang,~\IEEEmembership{Member,~IEEE}, \\Tianyao Huang,~\IEEEmembership{Member,~IEEE}, Feng Xu,~\IEEEmembership{Senior Member,~IEEE}, \\Yonina C. Eldar,~\IEEEmembership{Fellow,~IEEE}\\
	} 
	\thanks{
    Z. Liu and F. Xu are with the Key Lab for Information Science of Electromagnetic Wave (MoE), Fudan University, Shanghai 200433, China (e-mail: \{liuzy20; fengxu\}@fudan.edu.cn).
    
    H. Zhang is with the School of Communication and Information Engineering, Nanjing University of Posts and Telecommunications, Nanjing 210003, China (e-mail: haiyang.zhang@njupt.edu.cn).
    
    T. Huang is with the School of Computer and Communication Engineering, University of Science and Technology Beijing, Beijing, 100083, China (e-mail: huangtianyao@ustb.edu.cn).
    
    Y. C. Eldar is with the Faculty of Math and CS, Weizmann Institute of Science, Rehovot, Israel (e-mail: yonina.eldar@weizmann.ac.il).
    }
	
	\vspace{-1.0cm}
	
}
\vspace{-0.75cm}

\begin{document}
	
	\maketitle
	\pagestyle{plain}
	\thispagestyle{plain}
\begin{abstract}
\Ac{dfrc} technology is emerging in next-generation wireless systems. \Ac{ris} arrays have been suggested as a crucial sensor component of the \ac{dfrc}. 
In this paper, we propose a \ac{hris}-assisted \ac{mimo} \ac{dfrc} system, where the \ac{hris} is capable of reflecting communication signals to mobile users and receiving the scattering signal reflected from the radar target simultaneously. 
Under such a scenario, we are interested in characterizing the fundamental trade-off between radar sensing and communication. Specifically, we study the joint design of the beamforming vectors at the \ac{bs} and the parameter configuration of the \ac{hris} so as to maximize the  \ac{sinr} of the radar while guaranteeing a communication  \ac{sinr} requirement. 
To solve the formulated non-convex beamforming design problem, we propose an efficient \ac{ao} approach. In particular, for fixed beams at the \ac{bs}, we use a fast grid search-assisted auto gradient descent (FGS-AGD) algorithm to seek the best \ac{hris} configuration; Then, a closed-form \ac{bs} beamforming solution is obtained using semidefinite relaxation. Numerical results indicate that compared with benchmark schemes, the proposed approach is capable of improving the radar performance and communication quality significantly and simultaneously. 

\end{abstract}

\acresetall

\vspace{-0.4cm}
\section{Introduction}
	
\Ac{dfrc} systems have attracted significant attention in wireless networks. They enable radar sensing and communication functionalities by sharing the same hardware platform \cite{liu2022integrated}. \ac{dfrc} systems have the potential to be applied in autonomous driving, virtual reality, and other {control settings} due to their integrating radar and communication properties \cite{ma2020joint}. To enable the coexistence of both radar and communication, numerous researchers have studied the implementation of \ac{dfrc} in recent years. For example, the authors in \cite{huang2020majorcom} use the frequency and spatial agility properties of the carrier agile phased array radar to implement a \ac{dfrc} and achieved comparable communication performance to an independent device while guaranteeing radar performance. In \cite{hassanien2017dual}, the authors utilize frequency-hopping waveform codes to obtain different orthogonal radar waveforms and embed phase-shift keying to implement communication. Though \ac{dfrc} technology has made significant progress, it still faces several challenges in practice. A \ac{dfrc} system may suffer severe signal degradation when the user is sheltered by trees or buildings. In addition, when detecting short-range objects, the \ac{dfrc} antennas need to {operate} in duplex mode to avoid signal coupling between the transmit and receive channels, which is difficult to implement in most practical \acp{bs} \cite{sankar2021joint}.

Recently, \ac{ris} technology has emerged for future six-generation wireless systems. \ac{ris} is capable of enhancing communication performance by modifying the radio propagation, i.e., making the \ac{em} environment better for wireless communication \cite{elmossallamy2020reconfigurable,3}.   
 The passive reflection process of the \ac{ris} is similar to radar target scattering, which can be utilized to facilitate radar sensing by adjusting the reflection parameters of the \ac{ris} \cite{chepuri2022integrated}. 
Consequently, \ac{ris} has also been introduced to the field of \ac{dfrc} \cite{pan2021reconfigurable,pitilakis2020multi}, which enables broadening the communication \ac{fov} and radar detection by generating desirable reflecting beam patterns \cite{2}. 
However, the conventional \ac{ris}-assisted \ac{dfrc} systems have two drawbacks. First, due to the passive reflection properties of the \ac{ris}, using \ac{ris} to achieve communication will bring signal fading \cite{basar2021reconfigurable} and make transmitter-\ac{ris} and \ac{ris}-receiver links cascaded that are difficult to measure separately \cite{wang2021joint}. For instance, \ac{ris} transforms the coherent superposition from a \ac{bs} into an incoherent superposition of the \ac{bs}  transmitted signals and \ac{ris} reflected signals, and the original channel from \ac{bs} to users into \ac{bs}-\ac{ris} and \ac{ris}-user cascaded channel \cite{he2019cascaded}. The second drawback of the passive \ac{ris}-assisted \ac{dfrc} system is that one can only reflect the impinging signal and requires an additional radar receiver for radar echo acquisition. 

To overcome the aforementioned challenges with conventional \ac{ris} systems, \acp{hris} have been suggested, that combine \acp{ris} with \acp{dma} to provide a particular amount of energy to each reflection unit on the intelligent surface for signal reception \cite{alexandropoulos2021hybrid}. In contrast to conventional \ac{ris}-assisted systems, the \ac{hris} contains a few \ac{rf} connections to receive incoming signals while keeping the passive reflection performance of the meta-material elements.
    Recognizing the presence of various \acp{hris}, including the simultaneously transmitting and reflecting (STAR)-\ac{ris} discussed in \cite{ni2022star,wuchenyu2021,ahmed2023}, the Multi-Functional \ac{ris} (MF-\ac{ris}) referenced in \cite{wang2023mfris}, and the semi-passive \ac{ris} applied in \cite{ni2022star}, our system stands out due to its additional reception functionality.
    In our system, each element possesses the unique ability to both reflect and receive the incoming signal concurrently, setting it apart from existing \ac{hris} configurations. For convenience, we adopt the abbreviation '\ac{hris}' to represent our proposed \ac{hris} configuration in the subsequent sections.
An \ac{hris}-assisted system can employ \ac{hris}'s signals or combine \ac{hris} and \ac{bs} signals for signal processing in order to perform radar detection, sensing, positioning, and communication integration. Based on this incorporation of reflection and reception functionalities, \acp{hris} have potential applications in individual channel estimation \cite{zhang2022channel} and near-field user localization problems \cite{zhang2022hybrid}. 
Moreover, their potential to simultaneously enhance the performance of radar detection and user communication deserves exploring. This motivates the study of using the \ac{hris} as an efficient means of facilitating radar detection and user communication.

In this paper, we consider an \ac{hris}-assisted \ac{mimo} \ac{dfrc} system, where the \ac{hris} not only reflects the signal from the \ac{bs} to both the radar targets and communication users but also receives the echoes from the targets. The proposed \ac{hris}-assisted \ac{mimo} \ac{dfrc} system comprises a \ac{bs} and a \ac{hris} and the control center, which utilizes both \ac{bs} and \ac{hris} devices for sensing and communication. Based on the above architecture, the control center configures the transmitted signals of the \ac{bs}, and modifies the reflecting signal and receiving signal of the \ac{hris} to implement passing communication symbols to users and radar targets scattering echo reception. Compared with conventional \ac{mimo} \ac{dfrc}, the proposed system obtains improvements in both the radar and communication parts. Specifically, the presented \ac{dfrc} system broadens the view of communication and assists the communication with non-line of sight users  \cite{basar2019wireless,3}.
On the other hand, the radar receiver is embedded in the \ac{hris}, usually located remotely from the \ac{bs}, which does not require the \ac{bs} to work in full-duplex mode  \cite{zhang2022channel,alexandropoulos2021hybrid}. Moreover, the \ac{hris} is comprised of cheap meta-material elements and has low hardware complexity, which makes the \ac{hris}-assisted \ac{mimo} \ac{dfrc} system more portable and deployable \cite{alexandropoulos2020hardware}.

    To the best of our knowledge, we are the first to propose an \ac{hris}-assisted \ac{mimo} \ac{dfrc} system, which utilizes the unique capabilities of the \ac{hris} to enhance both radar and communication performance simultaneously. Differing from the existing STAR-\ac{ris} systems, the proposed system leverages the \ac{hris} to reflect the downlink communication signal to the user and receive the radar echo for sensing. Under such an \ac{hris}-assisted \ac{mimo} \ac{dfrc} system, our focus lies in the joint beamforming design at the \ac{hris} and \ac{bs} to achieve a balanced performance for both radar and communications. We formulate a comprehensive mathematical model for the \ac{hris}-assisted \ac{mimo} \ac{dfrc} system. Our model incorporates both the joint transmit beam signal processing carried out by the \ac{bs}, as well as the reflection and reception configuration designs of the \ac{hris}. Subsequently, a multi-parameter optimization problem corresponding to the joint beamforming of the \ac{hris}-assisted \ac{mimo} \ac{dfrc} system is formulated.

    Specifically, we jointly design the transmit beamforming vectors at the \ac{bs} and the configurations of \ac{hris} to maximize the radar performance while guaranteeing communication performance. However, the joint beamforming design is a multi-parameter and combinatorial optimization problem, which is non-convex and, consequently, challenging to solve. To address this, we propose an alternating method to solve the resulting problem by recasting the original problem into two sub-problems: the \ac{hris} configuration design and the design of the radar and communication transmitted beam of the \ac{bs}. For the solution of these sub-problems, we suggest an \ac{ao} approach to design the parameters of the \ac{hris} configuration and \ac{bs}'s beamforming vectors. In particular, a gradient descent-based algorithm is proposed for the \ac{hris} configuration design, and a \ac{sdr} technique with bisection search is applied for \ac{bs} beamforming optimization. However, due to the high complexity of the \ac{hris} configuration, we proposed a \ac{fgs} assisted \ac{agd} algorithm that combines the efficiency of fast grid search with the flexibility of PyTorch \ac{ad}. This approach is specifically designed to overcome the challenges of optimizing large \ac{hris} configurations.

    Extensive numerical results show that the performance of both radar and communication of \ac{hris}-assisted \ac{dfrc} outperforms the \acp{bs}-only and random-configured \ac{hris} systems. Furthermore, we study the impact of the integrated beampattern and power allocation of the \ac{hris} and the \acp{bs} on radar detection and wireless communication. In particular, under the same communication threshold and transmitted power, we show that the proposed system achieves a remarkable $3$~dB improvement in radar performance compared to the \acp{bs}-only system and an impressive $6$ dB gain over the random-configured \ac{hris} system.

The rest of this paper is organized as follows: Section \ref{sec:system} presents the \ac{hris}-assisted \ac{mimo} \ac{dfrc} system, reviews the considered antenna architectures, and formulates the joint optimization problem in our \ac{dfrc} system. Section \ref{sec:Solution} presents efficient methods to optimize the {beampattern} in both antenna architectures of the \ac{bs} and \ac{hris}, while Section \ref{sec:Sims} numerically demonstrates the solution of this proposed system and evaluates its performance in different \ac{dfrc} {settings}. Finally, Section \ref{sec:Conclusions} concludes the paper.

Throughout the paper, we use boldface lower-case and upper-case letters for vectors and matrices, respectively. For a matrix $\boldsymbol{A}$, the $(i, j)$-th element of $\boldsymbol{A}$
is denoted by $[\boldsymbol{A}]_{i,j}$. The $\ell_2$ norm, conjugate operation, transpose, Hermitian transpose, element-wise product, and stochastic expectation are written as $|\cdot |^2$, $(\cdot)^*$, $(\cdot)^T$, $(\cdot)^H$, $\odot$, and $\mathbf{E}( \cdot )$, respectively. {We use $\boldsymbol{I}_N$ to denote an $N$-dimensional
identity matrix, $\boldsymbol{0}_{M\times N}$ is an $M\times N$ zero matrix, and $\mathbbm{C}$ is the complex set.}

	\vspace{-0.2cm}
	\section{System Model and problem formulation}
	\label{sec:system}

In this section, we first present the general system model of the proposed \ac{hris}-assisted \ac{mimo} \ac{dfrc} system in Section~\ref{sec:system_model}. We describe \ac{hris} operation in Section~\ref{sec:pre_hris}. Then, we introduce the performance evaluation metrics of radar and communication functionalities in such a \ac{dfrc} system in Section~\ref{sec:PM_JRC}. Finally, we formulate the joint beamforming design problem in Section~\ref{sec:pb_formular}.

\vspace{-0.2cm}
\subsection{System Model}
\label{sec:system_model}

\begin{figure}[htbp]
    \centering
    \includegraphics[width=0.95\linewidth]{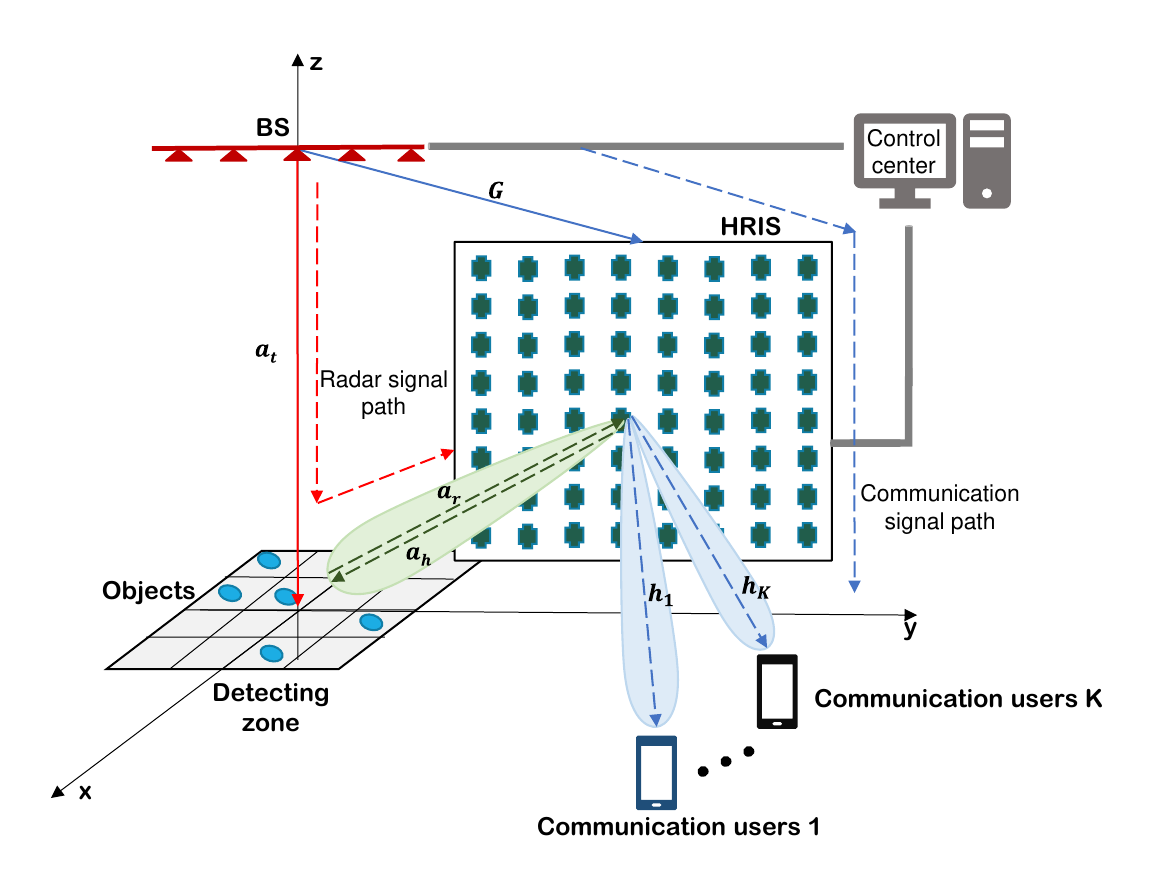}
    \caption{The geometry of detecting zone and user terminals.}
    \label{fig:f2}
\end{figure}

We consider an \ac{hris}-assisted \ac{mimo} \ac{dfrc} system consisting of a \ac{bs}, an \ac{hris}, detecting zone and communication user terminals. We assume the \ac{bs} consists of a \ac{ula} with $T$ antennas, and the \ac{hris} consists of $N$ elements. In this system, both the \ac{bs} and a \ac{hris} are connected to a data control center by cables and controlled by the control center. This setup is graphically presented in Fig. \ref{fig:f2}, and it incorporates the downlink communication and radar object detection with the assistance of an \ac{hris}. Specifically, the \ac{bs} sends both radar signals and communication signals to perform radar detection and communication simultaneously. As in \cite{zhang2022channel}, we assume there is no direct link between the \ac{bs} and communication users. The \ac{hris} helps forward the communication signals to the communication users and receive the radar signals reflected from the detecting zone.
In the case of multiple users, there are $K$ communication users who receive the communicated signals from the reflection of the \ac{hris} to communicate with the \ac{bs} independently. For radar detection, the detecting zone in the \ac{hris} system is the region where radar targets are located. The configuration of the detecting zone is determined by the antenna architecture of the \acp{bs}. To be specific, we divide the detecting zone into $P$ rows and $Q$ columns with the grid size being the field of view corresponding to the angle resolution of the \acp{bs}. Each block in the detecting zone can scatter the signal from both the \ac{bs} and the \ac{hris}, and the scattering signal is received by an \ac{rf} chain on the \ac{hris}.

We begin with the signals transmitted from the \ac{bs}. The discrete-time joint transmit beam signal of the \ac{bs} in time step $n$ is written as
\begin{equation}
\label{eq1}
    \begin{aligned}
    \centering
    \boldsymbol{x}(n)=\boldsymbol{W_cc}(n)+\boldsymbol{w}_{r}s(n), 
    \end{aligned}
\end{equation}
where $\boldsymbol{c}(n)=[c_1(n),...,c_K(n)]^T$ represents $K$ communication symbol streams intended for $K$ communication users, and $\boldsymbol{W}_c=[\boldsymbol{w}_1,...,\boldsymbol{w}_K]\in\mathbbm{C}^{T\times K}$ denotes the communication precoder matrix. Similarly, $s(n)$ is an individual radar waveform with unit power, and $\boldsymbol{w}_{r}\in\mathbbm{C}^{T\times 1}$ is the controllable radar beamforming vector of $T$ antennas. Without loss of generality, we assume each entry of the communication signals $\boldsymbol{c}(n)$ is a wide-sense stationary random process with zero-mean and unit power, {and uncorrelated with each other}, namely $\mathbf{E}(\boldsymbol{c}(n)\boldsymbol{c}^H(n))=\boldsymbol{I}_K$.
In addition, the communication symbols are uncorrelated with the radar waveform, meaning $\mathbf{E}(\boldsymbol{c}(n)s(n))=\boldsymbol{0}_{K\times1}$.

To illustrate the power constraint for the \ac{bs}, we denote the covariance matrix of the transmitted beamforming matrix by $\boldsymbol{R}\triangleq\boldsymbol{WW}^H$, where $\boldsymbol{W}=[\boldsymbol{W}_c,\boldsymbol{w}_{r}]\in \mathbbm{C}^{T\times (K+1)}$ is the joint controllable matrix of the \ac{bs}. Thus, each antenna's power constraint {implies} that
\begin{equation}
\label{eq_power}
    \begin{aligned}
    \centering
    [\boldsymbol{R}]_{j,j}=[\boldsymbol{W}_c\boldsymbol{W}_c^H + \boldsymbol{w}_{r}\boldsymbol{w}_{r}^H]_{j,j}=P_t,~j=1,...,T,
    \end{aligned}
\end{equation}
where $P_t$ is the transmit power for each antenna.

\vspace{-0.2cm}
\subsection{Preliminaries of HRIS}
\label{sec:pre_hris}
\ac{hris} is a new metamaterial device that achieves reflection and reception simultaneously. 
To achieve this hybrid operation, each metasurface element of the \ac{hris} must be capable of reflecting a component of the impinging signal while also receiving another portion of it in a controlled manner \cite{alexandropoulos2021hybrid}. The signals connected to the waveguides are then measured by the \ac{rf} chain and utilized to determine radar and communication information. In \cite{zhang2022channel}, the authors presented such a hybrid metamaterial surface {which} was applied to reflect in a reconfigurable function while using the received component of the signal to recover the target's angle of arrival locally. 
    Regarding the \ac{hris} protocol, \cite{Alexandropoulos2023hybrid} entirely covers the configurations of hybrid meta-atoms, implementation methodologies, full-wave \ac{em} simulations, and its application in 6G wireless technology.

We model the coexistence of reflection and reception functions with a hybrid metasurface composed of $N$ adjustable meta-atom elements.
One straightforward approach to implement this operation is to link individual elements to waveguides. Subsequently, the signals arrived at these waveguides are received by the attached \ac{rf} chain, which are then leveraged to extract essential information about radar sensing and communication. Meanwhile, these elements direct a portion of the impinging signal to the desired direction. 
\begin{figure}[htbp]
	\centering  
	\includegraphics[width=0.95\linewidth]{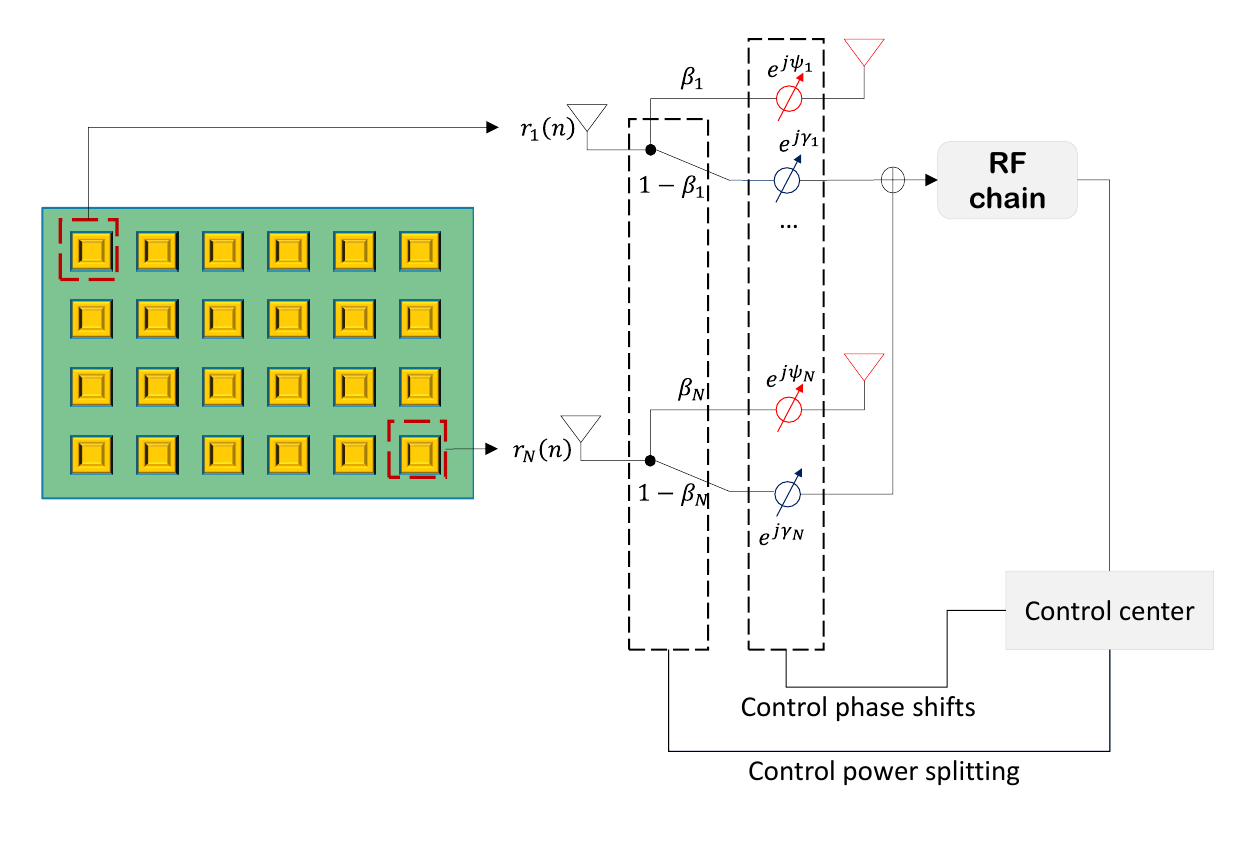}
	\caption{The receiving and reflecting operation of the \ac{hris}.}
	\label{function_HRIS}
\end{figure}
As illustrated in Fig. \ref{function_HRIS}, the control center can modify the reflected and received signals arriving at its surface by adjusting the surface's amplitude and phase shifts. Let $r_l(n)$ denote the discrete-time signal arriving at the $l$-th element of the \ac{hris} in time step $n$. Part of the signal is reflected to the desired direction with the adjustment by the parameter $\beta_l\in [0,1]$ and phase shift $\psi_l\in[0,2\pi)$. The forward reflected signal is consequently given as
\begin{equation}
\label{eqrf}
    \begin{aligned}
    \centering
    y_l^f(n)=\beta_le^{j\psi_l}r_l(n),~l=1,...,N.
    \end{aligned}
\end{equation}
Since per-element of the \ac{hris} enables to locally receive signals via analog combining and digital processing, the received signal collected by the \ac{rf} chain is expressed as
\begin{equation}
\label{eqrc}
    \begin{aligned}
    \centering
    y_{l}^r(n)=(1-\beta_l)e^{j\gamma_{l}}r_l(n),~l=1,...,N,
    \end{aligned}
\end{equation}
where $1-\beta_l$ is the amplitude allocated for receiving signal and $\gamma_{l}\in[0,2\pi)$ is an additional phase shift that controls the phaser connected to the \ac{rf} chain. 

By concatenating the arrived signal $r_l(n)$ and reflected signal $y_l^f(n)$ from the whole \ac{hris} into vectors $\boldsymbol{r}(n)$ and $\boldsymbol{y}^f(n)$, respectively, {the} received signal is formulated as
\begin{equation}
\label{eqrfh}
    \begin{aligned}
    \centering
    \boldsymbol{y}^f(n)=\boldsymbol{\Psi}(\boldsymbol{\beta},\boldsymbol{\psi})\boldsymbol{r}(n),
    \end{aligned}
\end{equation}
where the reflected matrix of the \ac{hris} is defined as $\boldsymbol{\Psi}(\boldsymbol{\beta},\boldsymbol{\psi})=\mathrm{diag}([\beta_1e^{j\psi_1},...,\beta_Ne^{j\psi_N}])$. Similarly, the reflected signal $y_{l}^r(n)$ from the \ac{rf} chain can be concatenated as a vector $\boldsymbol{y}^r(n)$ given by
\begin{equation}
\label{eqrch}
    \begin{aligned}
    \centering
    \boldsymbol{y}^r(n)=\boldsymbol{\phi}^H(\boldsymbol{\beta},\boldsymbol{\gamma})\boldsymbol{r}(n),
    \end{aligned}
\end{equation}
where the $l$-th element of the received vector is $[\boldsymbol{\phi}]_{l}=(1-\beta_l)e^{j\gamma_{l}}$.

The \ac{hris} enables to {change the \ac{em} environment} by externally controllable parameters. The beampattern adjustment performance is dominated by amplitude distribution on the surface of the \ac{hris} while slightly affected by the phased shifts in different elements of the \ac{hris} \cite{zhang2022channel}. Therefore, in this paper, we mainly control the beampattern of \ac{hris} by adjusting the power splitting factor $\boldsymbol{\beta}$
 with fixed phase shifts $\boldsymbol{\psi}$ and $\boldsymbol{\gamma}$. To be concrete, we optimize $\boldsymbol{\beta}$ with both $\boldsymbol{\psi}$ and $\boldsymbol{\gamma}$ set to zero in order to investigate the effect of power allocation on the \ac{hris}-assisted \ac{mimo} \ac{dfrc} system performance.

\vspace{-0.2cm}
\subsection{Performance Metrics of Radar and Communication}
\label{sec:PM_JRC}

    The main purpose of the \ac{hris}-assisted \ac{dfrc} system design is to achieve maximum radar sensing performance while ensuring efficient downlink multiuser communication. To meet multiuser communication requirements, both the \ac{bs} beamforming and \ac{hris} configuration are tailored. For radar detection tasks, \ac{bs}'s transmitted beam and \ac{hris}'s receiving beam are specifically designed toward the target located within the detecting zone. These performance metrics of multiuser communication and radar detection are formulated in the subsequent sections.

\subsubsection{The Evaluation Metric of Communication}

 Let $\boldsymbol{G}\in\mathbbm{C}^{N\times T}$ and $\boldsymbol{h}_k\in\mathbbm{C}^{N\times 1}$ represent the channel from \ac{bs} to \ac{hris} and the channel from the \ac{hris} to the $k$-th user, respectively. The received communication signal at the $k$-th user, denoted by $u_k(n)$,  can be written as
\begin{equation}
\label{eq4}
    \begin{aligned}
    \centering
    u_k(n) = \boldsymbol{h}_k^H\boldsymbol{\Psi}(\boldsymbol{\beta})\boldsymbol{Gx}(n) + v_k(n),
    \end{aligned}
\end{equation}
where $v_k(n)$ is \ac{awgn} with covariance $\sigma^2$.

We assemble the communication channels from the \ac{hris} to multiple users, which we assume to be known, in the complex matrix $\boldsymbol{H}\triangleq[\boldsymbol{h}_1,...,\boldsymbol{h}_K]^H\in\mathbbm{C}^{K\times N}$. The signal received at $K$ users is represented as the $K \times 1$ vector $\boldsymbol{u}(n)\triangleq[u_1(n),...,u_K(n)]^T$ and we collect the $v_k(n)$ into a $K\times1$ vector $\boldsymbol{v}(n)$. Combining (\ref{eq1}) and (\ref{eq4}), we rewrite the received signal at $K$ users into radar-to-users parts and inter-users parts, given by
\begin{equation}
\label{eq5}
    \begin{aligned}
    \centering
    \boldsymbol{u}(n)= \boldsymbol{H\Psi}(\boldsymbol{\beta})\boldsymbol{GW}_c\boldsymbol{c}(n)
    +\boldsymbol{H\Psi}(\boldsymbol{\beta})\boldsymbol{Gw}_{r}s(n)+\boldsymbol{v}(n).
    \end{aligned}
\end{equation}
Let the cascaded communication channel $\boldsymbol{H}_e=[\boldsymbol{\hat{h}}_1,...,\boldsymbol{\hat{h}}_K]^H\in\mathbbm{C}^{K\times T}$ be denoted as
\begin{equation}
\label{cascaded channel}
    \begin{aligned}
    \centering
    \boldsymbol{H}_e(\boldsymbol{\beta})=\boldsymbol{H\Psi}(\boldsymbol{\beta})\boldsymbol{G}.
    \end{aligned}
\end{equation}

Similar to  \cite{4}, we use the \ac{sinr} to evaluate communication performance. {The communication power of the $k$-th user is }
\begin{equation}
\label{user_power}
    \begin{aligned}
        \centering
        \mathbf{E}\left(|[\boldsymbol{H}_e(\boldsymbol{\beta})\boldsymbol{W}_c]_{k,k}c_k(n)|^2\right)=[\boldsymbol{H}_e(\boldsymbol{\beta})\boldsymbol{W}_c\boldsymbol{W}_c^H\boldsymbol{H}_e^H(\boldsymbol{\beta})]_{k,k}.
    \end{aligned}
\end{equation}
The communication inter-user interference power is given by
\begin{equation}
\label{eq9}
    \begin{aligned}
        \centering
        \mathbf{E}&\left(\sum_{i\ne k}^K|[\boldsymbol{H}_e(\boldsymbol{\beta})\boldsymbol{W}_c]_{k,i}c_k(n)|^2\right)\\
        &~~~~~~~~~~~~~~~~~~=\sum_{i\ne k}^K \left[\boldsymbol{H}_e(\boldsymbol{\beta})\boldsymbol{W}_c\boldsymbol{W}_c^H\boldsymbol{H}_e^H(\boldsymbol{\beta})\right]_{k,i}.
    \end{aligned}
\end{equation}
The interference power between the communication signals and the radar signal is
\begin{equation}
\label{eq_rc_in}
    \begin{aligned}
        \centering
        \mathbf{E}\left(|[\boldsymbol{H}_e(\boldsymbol{\beta})\boldsymbol{w}_{r}]_{k,1}s(n)|^2\right)=[\boldsymbol{H}_e(\boldsymbol{\beta})\boldsymbol{w}_{r}\boldsymbol{w}_{r}^H\boldsymbol{H}_e^H(\boldsymbol{\beta})]_{k,1}.
    \end{aligned}
\end{equation}
Therefore, the \ac{sinr} of the $k$-th user is expressed by (\ref{eq10}). 
\begin{figure*}
    \begin{equation}
    \label{eq10}
    \begin{aligned}
    \centering
    \eta_c(\boldsymbol{w}_{r},\boldsymbol{W}_c,\boldsymbol{\beta};k) = \frac{[\boldsymbol{H}_e(\boldsymbol{\beta})\boldsymbol{W}_c\boldsymbol{W}_c^H\boldsymbol{H}_e^H(\boldsymbol{\beta})]_{k,k}}{\sum_{i\ne k}^K [\boldsymbol{H}_e(\boldsymbol{\beta})\boldsymbol{W}_c\boldsymbol{W}_c^H\boldsymbol{H}_e^H(\boldsymbol{\beta})]_{k,i}+[\boldsymbol{H}_e(\boldsymbol{\beta})\boldsymbol{w}_{r}\boldsymbol{w}_{r}^H\boldsymbol{H}_e^H(\boldsymbol{\beta})]_{k,1}+\sigma^2}.
    \end{aligned}
\end{equation}
\hrulefill
\end{figure*}

For convenience, let each column of $\boldsymbol{W}$ which represents the controllable codes of the \ac{bs} be denoted by $\boldsymbol{w}_i$. The covariance matrix $\boldsymbol{R}$ of the transmitted beamforming matrix $\boldsymbol{W}$ can then be rewritten as the sum of sub-covariance matrices $\boldsymbol{R}_i\triangleq \boldsymbol{w}_i\boldsymbol{w}_i^H$ of different codes:
\begin{equation}
\label{eq22}
    \begin{aligned}
    \centering
    \boldsymbol{R}=\sum_{i=1}^{K+1}\boldsymbol{w}_i\boldsymbol{w}_i^H = \sum_{i=1}^{K+1}\boldsymbol{R}_i,
    \end{aligned}
\end{equation}
{where beamforming matrix $\boldsymbol{W}$ consists of $K$ communication precoder vectors and one radar precoder vector.} Combining with the definition of the $k$-th user's cascaded communication channel in (\ref{cascaded channel}) and substituting (\ref{eq22}) into (\ref{eq10}), the $k$-th user's \ac{sinr} can be expressed as (\ref{eq23}),
\begin{figure*}
    \begin{equation}
    \label{eq23}
    \begin{aligned}
    \centering
    \eta_c(\boldsymbol{R},\boldsymbol{R}_k, \boldsymbol{\beta}; k) = \frac{\boldsymbol{\hat{h}}_{k}^H(\boldsymbol{\beta})\boldsymbol{R}_k\boldsymbol{\hat{h}}_{k}(\boldsymbol{\beta})}{\boldsymbol{\hat{h}}_{k}^H(\boldsymbol{\beta})\boldsymbol{R}\boldsymbol{\hat{h}}_{k}(\boldsymbol{\beta})-\boldsymbol{\hat{h}}_{k}^H(\boldsymbol{\beta})\boldsymbol{R}_k\boldsymbol{\hat{h}}_{k}(\boldsymbol{\beta})+\sigma^2},~  k=1, ..., K.
    \end{aligned}
\end{equation}
\hrulefill
\end{figure*}
with the constraint (\ref{eq22}) and
\begin{equation}
\label{rank_1}
    \begin{aligned}
        \centering
        {\rm rank}(\boldsymbol{R}_i)=1,~ i=1,...,K+1.
    \end{aligned}
\end{equation}

\subsubsection{The Evaluation Metric of Radar Detection}

For radar detection, the \ac{bs} transmits the joint signals that illuminate both the detecting zone and the \ac{hris}. Subsequently, the \ac{hris} will reflect the incoming signal resulting from the \ac{bs}, directing it back towards the detecting zone. Therefore, the receiving signal at the \ac{hris} comprises two components.
    
The arriving signal from the \ac{bs} to the radar target can be expressed as
\begin{equation}
\label{eq12}
    \begin{aligned}
    \centering
    y(n;p,q)=\boldsymbol{a}_t^H(p,q)\boldsymbol{x}(n),
    \end{aligned}
\end{equation}
where $\boldsymbol{a}_t(p,q)\in\mathbbm{C}^{T\times 1}$ is the steering vector from the \ac{bs} to the $(p,q)$ block in the detecting zone. 
Combined with (\ref{eq4}), the signal forwarded by the \ac{hris} is
\begin{equation}
\label{eq13}
    \begin{aligned}
    \centering
    \boldsymbol{r}_e(n)=\boldsymbol{\Psi}(\boldsymbol{\beta})\boldsymbol{Gx}(n),
    \end{aligned}
\end{equation}
and the arriving signal from the \ac{hris} can be written as
\begin{equation}
\label{eq14}
    \begin{aligned}
    \centering
    r_s(n;p,q)={\boldsymbol a}_h^H(p,q)\boldsymbol{r}_e(n),
    \end{aligned}
\end{equation}
where $\boldsymbol{a}_h(p,q)\in\mathbbm{C}^{N\times 1}$ is the steering vector from the \ac{hris} to the $(p,q)$ block. 
Using the \ac{hris} to receive the scattering signal and perform radar detection, the received signal at the \ac{hris} can be consequently formulated as
\begin{equation}
\label{eq16}
    \begin{aligned}
    \centering
    \boldsymbol{r}(n;p,q)=\boldsymbol{\phi}^H(\boldsymbol{\beta})\boldsymbol{a}_r(p,q)(r_s(n;p,q)+y(n;p,q)),
    \end{aligned}
\end{equation}
where $\boldsymbol{a}_r(p,q)\in\mathbbm{C}^{N\times 1}$ is the steering vector from the $(p,q)$ block in the detecting zone to the $n$-th element on the \ac{hris}.

For convenience of the analysis, {we define the cascaded reflected vector $\boldsymbol{\hat{a}}_h\in\mathbbm{C}^{T\times 1}$ and the cascaded received scalar $A_r$ as 
\begin{subequations}
\label{cascaded reflect}
    \begin{align}
        \centering
        &\boldsymbol{\hat{a}}_h^H(\boldsymbol{\beta}) = \boldsymbol{a}_h^H(p,q)\boldsymbol{\Psi}(\boldsymbol{\beta})\boldsymbol{G}, \\
        & A_r(\boldsymbol{\beta})=\boldsymbol{\phi}^H(\boldsymbol{\beta})\boldsymbol{a}_r(p,q).
    \end{align}
\end{subequations}}
Similar to the work in \cite{pritzker2022transmit}, we use the \ac{sinr} as the radar performance metric. 
    To this end, we split the receiving signal at the \ac{rf} chain of the \ac{hris} into two components: the primary valid signal and the interference signal. In our considered \ac{hris}-assisted \ac{dfrc} system, the scattering signal that came from the illumination of the \ac{bs}'s transmitted waveform is assumed as the valid primary signal. Meanwhile, the scatter back resulting from the illumination of the \ac{hris}'s reflected beam is defined as interference.  
Then, by substituting the expressions of the reflected vector and received scalar into (\ref{eq16}), the valid signal is {$A_r(\boldsymbol{\beta})s(n)\boldsymbol{a}_t^H\boldsymbol{w}_{r}$}.
{The useful radar sensing power of the $(p,q)$ block is then derived as
\begin{equation}
\label{use_radar}
    \begin{aligned}
        \centering
        \mathbf{E}\left(|A_r(\boldsymbol{\beta})s(n)\boldsymbol{a}_t^H\boldsymbol{w}_{r}|^2\right)=|A_r(\boldsymbol{\beta})|^2\boldsymbol{a}_t^H \boldsymbol{w}_{r}\boldsymbol{w}_{r}^H\boldsymbol{a}_t.
    \end{aligned}
\end{equation}
The interference power from \ac{hris} to the $(p,q)$ block is
\begin{equation}
\label{inter_radar}
    \begin{aligned}
        \centering
        \mathbf{E}\left(|A_r(\boldsymbol{\beta})\boldsymbol{\hat{a}}_h^H(\boldsymbol{\beta})\boldsymbol{x}(n)|^2\right) = |A_r(\boldsymbol{\beta})|^2\boldsymbol{\hat{a}}_h^H(\boldsymbol{\beta})\boldsymbol{W}\boldsymbol{W}^H\boldsymbol{\hat{a}}_h(\boldsymbol{\beta}).
    \end{aligned}
\end{equation}
Combining (\ref{use_radar}) with (\ref{inter_radar}), the \ac{sinr} of the radar is expressed as
\begin{equation}
\label{eq18_}
    \begin{aligned}
    \centering
    \eta_r(\boldsymbol{W},\boldsymbol{\beta};p,q) =  \frac{|A_r(\boldsymbol{\beta})|^2\boldsymbol{a}_t^H \boldsymbol{w}_{r}\boldsymbol{w}_{r}^H\boldsymbol{a}_t}{|A_r(\boldsymbol{\beta})|^2\boldsymbol{\hat{a}}_h^H(\boldsymbol{\beta})\boldsymbol{W}\boldsymbol{W}^H\boldsymbol{\hat{a}}_h(\boldsymbol{\beta})+\sigma^2}.
    \end{aligned}
\end{equation}}
 
Furthermore, taking into account the scattering signal from the user, the expression of the \ac{sinr} for the radar is depicted as (\ref{expand_mdoel}),
    \begin{figure*}[ht]
    \begin{equation}
    \label{expand_mdoel}
        \begin{aligned}
            \centering
    \eta_r(\boldsymbol{W},\boldsymbol{\beta};p,q) =  \frac{|A_r(\boldsymbol{\beta})|^2\boldsymbol{a}_t^H \boldsymbol{w}_{r}\boldsymbol{w}_{r}^H\boldsymbol{a}_t}{|A_r(\boldsymbol{\beta})|^2\boldsymbol{\hat{a}}_h^H(\boldsymbol{\beta})\boldsymbol{W}\boldsymbol{W}^H\boldsymbol{\hat{a}}_h(\boldsymbol{\beta})+\boldsymbol{\phi}^H\boldsymbol{H}^H\boldsymbol{H}_e(\boldsymbol{\beta})\boldsymbol{WW}^H\boldsymbol{H}_e^H(\boldsymbol{\beta})\boldsymbol{H\phi}+\sigma^2},
        \end{aligned}
    \end{equation}
\hrulefill
\end{figure*}
    where $\boldsymbol{\phi}^H\boldsymbol{H}^H\boldsymbol{H}_e(\boldsymbol{\beta})\boldsymbol{WW}^H\boldsymbol{H}_e^H(\boldsymbol{\beta})\boldsymbol{H\phi}$ is the interference power from the user scattering. Our goal is to establish a foundational understanding of the \ac{hris}-assisted \ac{dfrc} system, focusing on the radar's \ac{sinr} without considering user scattering. In this work, we proceed to explore and evaluate the performance metric of radar via (\ref{eq18_}) in the following section. 

Compared with other \ac{dfrc} works \cite{4, pritzker2022transmit}, we use the same metrics to evaluate radar and communication performance. However, due to the coexistence of the \ac{hris} device and the \ac{bs} in our proposed \ac{dfrc} system, we have another parameter $\boldsymbol{\beta}$ that affects both radar and communication performance.

\vspace{-0.2cm}
\subsection{Problem formulation}
\label{sec:pb_formular}
From (\ref{eq23}) and (\ref{eq18_}), both the radar and communication performance evaluation metrics are a function of the controllable coefficients in the \ac{bs} and the \ac{hris} {which are the beamforming matrix $\boldsymbol{W}=[\boldsymbol{w}_1,...,\boldsymbol{w}_{K+1}]$ and the power splitting factor $\boldsymbol{\beta}$.} As the \ac{hris}-assisted \ac{mimo} \ac{dfrc} system has the same controllable coefficients to determine the radar and communication performance, there is a trade-off between them.
Here, we are interested in characterizing this trade-off by concurrently optimizing the joint beamforming matrix of the \ac{bs} and configuring the parameters of the \ac{hris}. To this end, we aim at maximizing the radar \ac{sinr} $\eta_r$ while guaranteeing communication performance $\Gamma_c$. 
The resulting problem can be formulated as a joint beamforming optimization problem:
{\begin{subequations}
\label{eq19}
    \begin{align}
    \centering
    \max_{\boldsymbol{W},\boldsymbol{\beta}}& \  \eta_r(\boldsymbol{W},\boldsymbol{\beta};p,q),\\
    {\rm s.t.} \ &\eta_c(\boldsymbol{R},\boldsymbol{R}_k, \boldsymbol{\beta}; k)\ge \Gamma_c,~ k=1,...,K , \label{c_cons1}\\
    & \boldsymbol{R}_i = \boldsymbol{w}_i\boldsymbol{w}_i^H,~i=1,...,K+1,\\
    & \boldsymbol{R}=\sum_{i=1}^{K+1}\boldsymbol{R}_i,\label{R_cov}\\
    & 0\leq\beta_l\leq 1 , ~l=1,...,N, \label{h_poewr}\\
    & [\boldsymbol{R}]_{j,j}=P_t, ~j=1,...,T \label{b_power},
    \end{align}
\end{subequations}
where $\boldsymbol{w}_i$ is the $i$-th column of the $\boldsymbol{W}$, $\eta_r(\boldsymbol{W},\boldsymbol{\beta};p,q)$ and $\eta_c(\boldsymbol{R},\boldsymbol{R}_k, \boldsymbol{\beta}; k)$ are defined by (\ref{eq18_}) and (\ref{eq23}), respectively.}


The objective of (\ref{eq19}) is the radar performance of the $\left(p,q\right)$ block detecting zone that we are interested in; (\ref{c_cons1})  is the communication constraint, with $\Gamma_c$ denoting the minimum communication \ac{sinr} requirement. 
Specifically, (\ref{R_cov}) is defined in (\ref{eq22}) and explains the covariance matrix $\boldsymbol{R}$ of the \ac{bs} transmitted beamforming signal is the sum of sub-covariance matrices $\boldsymbol{R}_i$ of different codes.
For the power constraints, (\ref{h_poewr}) denotes the power allocation of each element in the \ac{hris}, and (\ref{b_power}) is the antenna power budget on the \ac{bs}.

Since the optimized parameters controlling the configuration of the \ac{bs} and the \ac{hris} are coupled, the problem formulated in (\ref{eq19}) is non-convex. Moreover, compared to \cite{4, pritzker2022transmit}, we need to optimize the beamforming matrices of the \ac{bs} and the power splitting factor $\boldsymbol{\beta}$ of the \ac{hris} simultaneously. To address this challenging problem, we propose an efficient \ac{ao} algorithm, which will be detailed in the next section.

\vspace{-0.2cm}
\section{Proposed alternating optimization algorithm}
\label{sec:Solution}
	
In this section, we develop an \ac{ao} algorithm to optimize the beamforming matrices of the \ac{bs} and \ac{hris} configuration parameters.  We begin with introducing the \ac{hris} optimization for fixed \ac{bs} beamforming matrices in Section~\ref{sec:hris-design}, which we then utilize to design the \ac{bs} beamforming matrices in Section~\ref{sec:bs design}. The parameters update strategy during each iteration is summarized in Section~\ref{sec:alt_beam}.

\vspace{-0.2cm}
\subsection{HRIS configuration with fixed BS beamforming matrices }\label{sec:hris-design}	


In this section, we optimize the \ac{hris} configuration with fixed  \ac{bs} beamforming matrix. Let $\boldsymbol{\bar{W}}^{(t)}=[\boldsymbol{\bar{W}}_{c}^{(t)},\boldsymbol{\bar{w}}_{r}^{(t)}]$ denote the fixed beamforming matrices at the \acp{bs} in the  $t$-th iteration, with $\boldsymbol{\bar{W}}_{c}^{(t)}=[\boldsymbol{\bar{w}}_{1}^{(t)},...,\boldsymbol{\bar{w}}_{K}^{(t)}]$ and $\boldsymbol{\bar{w}}_{r}^{(t)}$ denoting the corresponding communication precoder matrix and radar beamforming vector, respectively. In this case, (\ref{eq19}) can be simplified as
{\begin{subequations}
\label{eq19h}
    \begin{align}
    \centering
    \max_{\boldsymbol{\beta}}&\  \eta_r(\boldsymbol{\bar{W}}^{(t)},\boldsymbol{\beta};p,q),\\
    {\rm s.t.} \ &\eta_c(\boldsymbol{\bar{R}}^{(t)},\boldsymbol{\bar{R}}_{k}^{(t)},\boldsymbol{\beta};k)\ge \Gamma_c,~ k=1,...,K ,\\
    & \boldsymbol{\bar{R}}_{i}^{(t)}=\boldsymbol{\bar{w}}_{i}^{(t)}\boldsymbol{\bar{w}}_{i}^{(t)H},~i=1,...,K+1,\\
    &
    \boldsymbol{\bar{R}}^{(t)}=\sum_{i=1}^{K+1}\boldsymbol{\bar{R}}_i^{(t)},\\
    & 0\leq\beta_l\leq1 , ~l=1,...,N,
    \end{align}
\end{subequations}}
where $\boldsymbol{\bar{R}}^{(t)}$ and $\boldsymbol{\bar{R}}_i^{(t)}$ are the covariance matrix and the $i$-th sub-covariance matrix in the $t$-th iteration, respectively.
We next express $\eta_c(\boldsymbol{\bar{R}}^{(t)},\boldsymbol{\bar{R}}_{k}^{(t)},\boldsymbol{\beta};k)$ and $\eta_{r}(\boldsymbol{\bar{W}}^{(t)},\boldsymbol{\beta};p,q)$ as functions of the power splitting factor $\boldsymbol{\beta}$ in the following proposition.

\begin{proposition}
\label{proposition1}
Given the noise variance $\sigma^2$, {the channel from \ac{bs} to \ac{hris} $\boldsymbol{G}$, and the channel from the \ac{hris} to the $k$-th user $\boldsymbol{h}_k$,} the communication user's \ac{sinr} and the radar \ac{sinr} can be recast as
\begin{subequations}
\label{eta_r_c}
    \begin{align}
    \centering
    &\eta_c(\boldsymbol{\beta};k) = \frac{\boldsymbol{\beta}^T\boldsymbol{C}_3\boldsymbol{\beta}}{\boldsymbol{\hat{h}}_k^H\boldsymbol{\bar{R}}^{(t)}\boldsymbol{\hat{h}}_k-\boldsymbol{\beta}^T\boldsymbol{C}_3\boldsymbol{\beta}+\sigma^2}, ~k=1,...,K, \label{eta_c}\\
    &\eta_r(\boldsymbol{\beta};p,q) = \frac{(1- \boldsymbol{\beta}^T)\boldsymbol{C}_1(1-\boldsymbol{\beta})}{(1-\boldsymbol{\beta}^T)\boldsymbol{a}_r\boldsymbol{\beta}^T \boldsymbol{C}_2\boldsymbol{\beta}\boldsymbol{a}_r^H(1-\boldsymbol{\beta})+\sigma^2},\label{eta_r}
    \end{align}
\end{subequations}
where $\boldsymbol{\hat{h}}_k$ is the cascaded communication channel calculated by (\ref{cascaded channel}), $\boldsymbol{C}_1$,  $\boldsymbol{C}_2$,  $\boldsymbol{C}_3$ are Hermitian matrices defined as follows:
\begin{subequations}
\label{C_1_2_3}
    \begin{align}
    \centering
    & \boldsymbol{C}_1 \triangleq \left(\boldsymbol{a}_t^H \boldsymbol{\bar{w}}_{r}^{(t)}\boldsymbol{\bar{w}}_{r}^{(t)H}\boldsymbol{a}_t\right)\boldsymbol{a}_r\boldsymbol{a}_r^H,\\
    & \boldsymbol{C}_2 \triangleq \boldsymbol{a}_h\odot\left(\boldsymbol{G}\boldsymbol{\bar{W}}^{(t)}\right)\left(\boldsymbol{a}_h\odot\left(\boldsymbol{G}\boldsymbol{\bar{W}}^{(t)}\right)\right)^H,\\
    & \boldsymbol{C}_3 \triangleq \boldsymbol{h}_k\odot\left(\boldsymbol{G}\boldsymbol{\bar{w}}_{k}^{(t)}\right)\left(\boldsymbol{h}_k\odot\left(\boldsymbol{G}\boldsymbol{\bar{w}}_{k}^{(t)}\right)\right)^H,~k=1,..,K.
    \end{align}
\end{subequations}
{Here, $\boldsymbol{\bar{W}}^{(t)}=\left[\boldsymbol{\bar{w}}_1^{(t)},...,\boldsymbol{\bar{w}}_K^{(t)},\boldsymbol{\bar{w}}_{r}^{(t)}\right]$ is the fixed \ac{bs}'s beamforming matrix for communication and radar sensing, $\boldsymbol{a}_t$ is the given transmitted steering vector of radar, and $\boldsymbol{a}_h$ and $\boldsymbol{a}_r$ are the given reflection and reception steering vectors of \ac{hris}, respectively.}
\end{proposition}

\begin{IEEEproof}
See Appendix \ref{app:Proof_1}.
\end{IEEEproof}

Using Proposition \ref{proposition1}, the \ac{hris} configuration design problem \eqref{eq19h} can be reformulated as
\begin{subequations}
\label{obj}
    \begin{align}
    \centering
    \max_{\boldsymbol{\beta}} \quad & \frac{(1- \boldsymbol{\beta}^T)\boldsymbol{C}_1(1-\boldsymbol{\beta})}{(1-\boldsymbol{\beta}^T)\boldsymbol{a}_r\boldsymbol{\beta}^T \boldsymbol{C}_2\boldsymbol{\beta}\boldsymbol{a}_r^H(1-\boldsymbol{\beta})+\sigma^2},\\
    {\rm s.t.} \ & {\frac{\boldsymbol{\beta}^T\boldsymbol{C}_3\boldsymbol{\beta}}{\boldsymbol{\hat{h}}_k^H\boldsymbol{\bar{R}}^{(t)}\boldsymbol{\hat{h}}_k-\boldsymbol{\beta}^T\boldsymbol{C}_3\boldsymbol{\beta}+\sigma^2}\geq \Gamma_c },~k=1,...,K,\label{com_beta}\\
    & 0\leq\beta_l\leq 1 , ~l=1,...,N.
    \end{align}
\end{subequations}
Problem (\ref{obj}) is still non-convex and challenging to solve.
To proceed, we consider the following optimization problem:
\begin{equation}
\label{objlag}
    \begin{aligned}
    \min_{\boldsymbol{\beta}\in \mathcal{S}}f(\boldsymbol{\beta}) \triangleq - \eta_r(\boldsymbol{\beta};p,q)
    + \mathcal{R}_\mathcal{S}(\boldsymbol{\beta}),
    \end{aligned}
\end{equation}
where $\mathcal{S}$ represents the feasible set of the \ac{hris}'s power splitting vector $\boldsymbol{\beta}$, given by $\mathcal{S}\triangleq \{\boldsymbol{\beta}\in \mathbbm{R}:0\leq \beta_l\leq1,~l=1,...,N \}$, and $\mathcal{R}_\mathcal{S}(\boldsymbol{\beta})$ is a Lagrangian operator, which converts the communication constraint (\ref{com_beta}) $\eta_c(\boldsymbol{\beta};k)\geq\Gamma_c$, for $k=1,..,K$ into the objective function.
To this end, we apply the Lagrangian operator:
\begin{equation}
\label{lag}    
    \begin{aligned}
        \centering
        \mathcal{R}_\mathcal{S}(\boldsymbol{\beta})=\lambda_1\cdot\underbrace{ \alpha_1^{\Gamma_c - \eta_c(\boldsymbol{\beta};k)}}_{\triangleq g_1(\boldsymbol{\beta})}
    + \lambda_2\cdot\sum_{l=1}^{N}\underbrace{(2\beta_l-1)^{\alpha_2}}_{\triangleq g_2({\beta}_l)}, 
    \end{aligned}
\end{equation}
where $\lambda_1$ and $\lambda_2$ are strictly positive hyper-parameters to control the threshold of the communication user's \ac{sinr} and manage the power of the \ac{hris}. Formally, the communication guarantee and power constraints of the \ac{hris} in (\ref{obj}) are converted into $g_1(\boldsymbol{\beta})$ and $g_2({\beta}_l)$, respectively. Combining the properties of the exponential and power functions, {$g_1(\boldsymbol{\beta})$ is designed as the exponential function which needs to satisfy:~$0<g_1(\boldsymbol{\beta})\leq 1$, when $\eta_c(\boldsymbol{\beta};k)\geq \Gamma_c,~\forall k\in[1,K]$; $g_2({\beta}_l)$ is considered as the even power function to add a high penalty to the point approaching the boundary of $\mathcal{S}$: $g_2(\beta_l)\gg 1$, when $\beta_l\notin [0,1]$. Here, we set $\alpha_1=4$ and $\alpha_2=10$ empirically.}

We next propose a \acf{fgs} assisted \acf{agd} algorithm to address the non-convex unconstrained optimization problem in (\ref{objlag}). The key idea of the proposed algorithm is to utilize \ac{ad} \cite{baydin2018automatic} to obtain a convergent solution according to the specific initialization parameters $\boldsymbol{\beta}$ generated by the proposed \ac{fgs}.

We begin with the implementation of \ac{agd}, and then introduce the designed initialization strategy \ac{fgs}.
First, to reduce the complexity of gradient calculation, we employ the advanced \ac{agd} approach {to obtain the solution of $\boldsymbol{\beta}$ after $M$ iterations}, as shown in Algorithm~\ref{al1_agd}. The $(i+1)$-th iterative step in conventional gradient descent is written as:
\begin{equation}
    \begin{aligned}
    \centering
    \boldsymbol{\beta}^{i+1} = \boldsymbol{\beta}^i + \Delta_i\cdot\nabla f(\boldsymbol{\beta}^i),
    \end{aligned}
\end{equation}
where $\nabla(\cdot)$ is the first-order gradient operator, and $\Delta_i$ is the step size.
{We replace the traditional gradient operator $\nabla(\cdot)$ with the \ac{ad} tool. The main concept of \ac{ad} is to represent the objective function as a computational graph according to chain rules, to which the back-propagation algorithm is applied.} Thus, the gradient direction at point $\boldsymbol{\beta}^i$ is given by 
\begin{equation}
\label{grad_dire}
    \begin{aligned}
        \centering
        \nabla f(\boldsymbol{\beta}^i) \approx \mathbf{BP}(f(\boldsymbol{\beta}^i)),
    \end{aligned}
\end{equation}
where $\mathbf{BP}(\cdot)$ is the gradient computation operator based on \ac{ad} and $f(\boldsymbol{\beta}^i)$ is the input objective function. Here, we use autograd of PyTorch \cite{paszke2019pytorch} to {compute} $\mathbf{BP}(\cdot)$ {and generate the initial $\boldsymbol{\beta}^0$ randomly}.
The gradient descent's step length $\Delta_i$ is substituted by the learning rate $l_r$ and is updated by the Adam optimizer of the PyTorch \cite{kingma2014adam}. Therefore, we can achieve \ac{agd} with variable step size by adjusting the learning rate. 

\begin{algorithm}
    \caption{\ac{hris} configuration via \ac{agd} algorithm}
    \renewcommand{\algorithmicrequire}{\textbf{Input:}}
    \renewcommand{\algorithmicensure}{\textbf{Output:}}
    \label{al1_agd}
    \begin{algorithmic}[1]
        \REQUIRE \leavevmode \\
        Initialize: $\boldsymbol{\beta }^0$, $M$, $l_r$;\\
        \FOR{$i=0:M-1$}
        \STATE Calculate the objective $f(\boldsymbol{\beta}^i) = - \eta_r(\boldsymbol{\beta}^i;p,q)+\mathcal{R}_\mathcal{S}(\boldsymbol{\beta}^i)$;
        \STATE Calculate the gradient direction $\nabla f(\boldsymbol{\beta}^i)$ according to (\ref{grad_dire});
        \STATE Find the next point and update $\boldsymbol{\beta}^{i+1} = \boldsymbol{\beta}^i+ l_r\cdot\nabla f(\boldsymbol{\beta}^i)$;
        \ENDFOR
        \ENSURE \leavevmode \\
        \ac{hris} configuration vector $\boldsymbol{\beta}$.
    \end{algorithmic}
\end{algorithm}

{The performance of Algorithm~\ref{al1_agd} depends heavily on the initial point $\boldsymbol{\beta}^0$. 
We design an \ac{fgs} algorithm to find a good initial $\boldsymbol{\beta}$, as shown in Algorithm \ref{al1}. 
    The fundamental principle of \ac{fgs} is to set the value corresponding to the first half entries of the vector $\boldsymbol{\beta}$ to zero::~$\boldsymbol{\beta}(1:\frac{N}{2})=\boldsymbol{0}_{\frac{N}{2}\times 1}$. Subsequently, the latter half of $\boldsymbol{\beta}$ is set as the unit column vector, following the multiplying of a coefficient varying in the space $\mathcal{S}$:~$\boldsymbol{\beta}(\frac{N}{2}+1:N)=z\Delta_z \boldsymbol{I}_{\frac{N}{2}\times 1}$, for $z=0,...,Z_{\max}$.
We define the number of grids in space $\mathcal{S}$ by $Z_{\max}$ and the grid step by $\Delta_z$, which satisfies:~$Z_{\max}\Delta_z=1$. Since the gird value $z\Delta_z$ of the slice of $\boldsymbol{\beta}$ is discrete and finite, the optimal solution of (\ref{objlag}) is obtained by traversing all grid values in space $\mathcal{S}$. In our designed \ac{fgs} algorithm, $\boldsymbol{\beta}$ is divided by $M_{\max}$ times to increase the freedom of the grid search. In the $m$-th iteration, we update the slice $(\zeta_1^m:\zeta_2^m)$ of the $\boldsymbol{\beta}^{m}$ by:
\begin{subequations}
\label{slices}
    \begin{align}
        \centering
        & (\zeta_1^m:\zeta_2^m)=\left(1:\frac{N}{2^m}\right),\\
        & (\zeta_1^m:\zeta_2^m)=\left(\frac{2^m-1}{2^m}N:N\right),
    \end{align}
\end{subequations}
while holding the same search space, expressed by
\begin{equation}
\label{fgs_grid}
    \begin{aligned}
        \centering
        \boldsymbol{\beta}^{m,z}(\zeta_1^m:\zeta_2^m)=z\Delta_z \boldsymbol{I}_{(\zeta_2^m-\zeta_1^m)\times 1},~ z= 0,...,Z_{\max},
    \end{aligned}
\end{equation}
Since the length of each slice needs to be not less than 1, the maximum iteration step size is $M_{\max}=\log_2^N$. 
Following the above discretization method, there exists an optimal solution $\boldsymbol{\beta}^{m}$ piloting to the minimal value of (\ref{objlag}). The $\boldsymbol{\beta}^m$ is thus updated based on the following operation:
\begin{equation}
\label{fgs_grid_up}
    \begin{aligned}
        \centering
        \boldsymbol{\beta}^{m+1} = \min_{\boldsymbol{\beta}^{m,z}} f\left(\boldsymbol{\beta}^{m,z}\right), ~ z= 0,...,Z_{\max}.
    \end{aligned}
\end{equation}}

\begin{algorithm}
    \caption{Initialization operation via \ac{fgs} algorithm}
    \renewcommand{\algorithmicrequire}{\textbf{Input:}}
    \renewcommand{\algorithmicensure}{\textbf{Output:}}
    \label{al1}
    \begin{algorithmic}[1]
        \REQUIRE \leavevmode \\
        Initialize: $\boldsymbol{\beta}^0=\boldsymbol{0}_{N\times1}$;\\
        \FOR{$m=1,...,M_{\max}$}
        \STATE Select the update slice $(\zeta_1^m:\zeta_2^m)$ {via (\ref{slices})};
        \FOR{$z=1,...,Z_{\max}$}
        \STATE Grid search $\boldsymbol{\beta}^{m,z}$ according to (\ref{fgs_grid});
        \STATE Calculate the objective function $f\left(\boldsymbol{\beta}^{m,z}\right)$;
        \ENDFOR
        \STATE Update the next point $\boldsymbol{\beta}^{m+1}$ according to (\ref{fgs_grid_up});
        \ENDFOR
        \ENSURE \leavevmode \\
        \ac{hris} initial point $\boldsymbol{\beta}^0$ used in Algorithm~\ref{al1_agd}.
    \end{algorithmic}
\end{algorithm}

In summary, we first proposed the \ac{fgs} approach to obtain a good initial point $\boldsymbol{\beta}^0$, which determines the boundary of the gradient descent. Then, the \ac{agd} approach based on the back-propagation of PyTorch's autograd is applied to find a solution of (\ref{objlag}).

\vspace{-0.2cm}
\subsection{Transmitted Beam Design with fixed HRIS configuration}
\label{sec:bs design}
\vspace{-0.1cm} 
After the \ac{hris} configuration optimization, the power splitting vector $\boldsymbol{\bar{\beta}}^{(t)}$ of the reflected matrix $\boldsymbol{\Psi}$ and received vector $\boldsymbol{\phi}$ are obtained. We henceforth seek to optimize the joint beamforming matrix $\boldsymbol{W}=[\boldsymbol{W}_c,\boldsymbol{w}_{r}]$ of the \ac{bs} to maximize the radar performance with the fixed \ac{hris} configuration. The optimization problem (\ref{eq19}) becomes: 
{\begin{subequations}
\label{eq19b}
    \begin{align}
    \centering
    \max_{\boldsymbol{W}}& \  \eta_r\left(\boldsymbol{W},\boldsymbol{\bar{\beta}}^{(t)};p,q\right),\\
    {\rm s.t.} \ &\eta_c\left(\boldsymbol{R},\boldsymbol{R}_k, \boldsymbol{\bar{\beta}}^{(t)}; k\right)\ge \Gamma_c, ~k=1,...,K ,\\
    & \boldsymbol{R}_i = \boldsymbol{w}_i\boldsymbol{w}_i^H,~i=1,...,K+1,\\
    & \boldsymbol{R}= \sum_{i=1}^{K+1}\boldsymbol{R}_i,\\
    & [\boldsymbol{R}]_{j,j}=P_t, ~j=1,...,T.
    \end{align}
\end{subequations}}

{Similar to the expression of the user's \ac{sinr}, we represent the radar's \ac{sinr} in terms of the transmit sub-covariance matrices $\boldsymbol{ R}_i$
\begin{equation}
\label{eq18}
    \begin{aligned}
    \centering
    \eta_r(\boldsymbol{R},\boldsymbol{R}_{K+1}, \boldsymbol{\beta};p,q) =  \frac{|A_r(\boldsymbol{\beta})|^2\boldsymbol{a}_t^H \boldsymbol{R}_{K+1}\boldsymbol{a}_t}{|A_r(\boldsymbol{\beta})|^2\boldsymbol{\hat{a}}_h^H(\boldsymbol{\beta})\boldsymbol{R}\boldsymbol{\hat{a}}_h(\boldsymbol{\beta})+\sigma^2},
    \end{aligned}
\end{equation}
where $\boldsymbol{R}_{K+1}=\boldsymbol{w}_r\boldsymbol{w}_r^H$.
By updating the cascaded received scalar $\bar{A}_{r}^{(t)}$ and the cascaded reflected vector $\boldsymbol{\bar{a}}_{h}^{(t)}$ based on (\ref{cascaded reflect}) and replacing $\boldsymbol{R}_{K+1}=\left(\boldsymbol{R}-\sum_{k=1}^K\boldsymbol{R}_{k}\right)$ according to (\ref{eq22}),} the \ac{sinr} of the radar for a point target at position $(p,q)$ is
\begin{equation}
\label{eq33}
    \begin{aligned}
    \centering
    \eta_r(\boldsymbol{R}, \boldsymbol{R}_{k};p,q)=\frac{\left|\bar{A}_{r}^{(t)}\right|^2\boldsymbol{ a}_t^H\left(\boldsymbol{R}-\sum_{k=1}^K\boldsymbol{R}_{k}\right)\boldsymbol{a}_t}{\left|\bar{A}_{r}^{(t)}\right|^2\boldsymbol{\bar{a}}_{h}^{(t)H}\boldsymbol{R}\boldsymbol{\bar{a}}_{h}^{(t)}+\sigma^2}.
    \end{aligned}
\end{equation}

Thus, with the respective expression of radar's \ac{sinr} $\eta_r$ and communication's \ac{sinr} $\eta_c$ in (\ref{eq33}) and (\ref{eq23}) in terms of $\boldsymbol{R}_i$, (\ref{eq19b}) is reformulated as
\begin{subequations}
\label{st2}
    \begin{align}
    \centering
    \max_{\boldsymbol{R}_1,...,\boldsymbol{R}_K,\boldsymbol{R}_{K+1}} \quad & \eta_r(\boldsymbol{R}, \boldsymbol{R}_{k};p,q),\\
    {\rm s.t.} \
    & \eta_c(\boldsymbol{R},\boldsymbol{R}_k; k)\ge\Gamma_c,~ k=1,...,K ,\\
    & \boldsymbol{R}=\sum_{i=1}^{K+1}\boldsymbol{R}_i,\\
    & {\rm rank}(\boldsymbol{R}_i)=1,~ i=1,...,K+1,\\
    & [\boldsymbol{R}]_{j, j}=P_t, ~j=1,...,T.
    \end{align}
\end{subequations}
 
The \ac{bs} transmitted beam design problem (\ref{st2}) is non-convex. 
According to \cite{zhang2010relationship}, (\ref{st2}) can be reduced to solving a sequence of convex feasibility problems. Thus, (\ref{st2}) is subsequently transformed as
\begin{subequations}
\label{st3}
    \begin{align}
    \centering
    \max_{\boldsymbol{R}_1,...,\boldsymbol{R}_K,\boldsymbol{R}_{K+1},\Gamma_r} ~&\Gamma_r,\\
    {\rm s.t.} \ & \eta_r(\boldsymbol{R}, \boldsymbol{R}_{k};p,q)\ge\Gamma_r,\\
    &\eta_c(\boldsymbol{R},\boldsymbol{R}_{k};k)\ge\Gamma_c,~ k=1,...,K ,\\
    & \boldsymbol{R}=\sum_{i=1}^{K+1}\boldsymbol{R}_i,\\
    & {\rm rank}(\boldsymbol{R}_i)=1,~ i=1,...,K+1,\\
    & [\boldsymbol{R}]_{j,j}=P_t, ~j=1,...,T.
    \end{align}
\end{subequations}

Let $\Gamma_r^*$ be the optimal solution of problem (\ref{st3}). Obviously, $\Gamma_r^*$ is also the optimal value of the original problem (\ref{st2}). If the following feasibility problem
\begin{subequations}
\label{st3_}
    \begin{align}
    \centering
    {\rm Find} ~&\boldsymbol{R}_1,...,\boldsymbol{R}_K,\boldsymbol{R}_{K+1},\\
    {\rm s.t.} \ & \eta_r(\boldsymbol{R}, \boldsymbol{R}_{k};p,q)\ge\Gamma_r,\\
    &\eta_c(\boldsymbol{R},\boldsymbol{R}_{k};k)\ge\Gamma_c,~ k=1,...,K ,\\
    & \boldsymbol{R}=\sum_{i=1}^{K+1}\boldsymbol{R}_i,\\
    & {\rm rank}(\boldsymbol{R}_i)=1,~ i=1,...,K+1,\\
    & [\boldsymbol{R}]_{j,j}=P_t, ~j=1,...,T,
    \end{align}
\end{subequations}
for a fixed $\Gamma_r$ is feasible, then it follows that $\Gamma_r^*\ge\Gamma_r$. If (\ref{st3_}) is infeasible, then $\Gamma_r^*<\Gamma_r$. Therefore, by giving a potential range of $\Gamma_r$ which contains $\Gamma_r^*$, the optimal solution of (\ref{st3}) can be obtained via bisection search. 

We next introduce the solution of (\ref{st3_}) which is still non-convex because of the rank-one constraints. It was shown in previous works \cite{4,pritzker2022transmit} that (\ref{st3_}) can be solved with \ac{sdr} without loss of optimality.
Omitting the rank-one constraint, (\ref{st3_}) is relaxed to
\begin{subequations}
\label{st5}
    \begin{align}
    \centering
    {\rm Find} ~&\boldsymbol{R}_1,...,\boldsymbol{R}_K,\boldsymbol{R}_{K+1},\\
    {\rm s.t.} \ & \eta_r(\boldsymbol{R}, \boldsymbol{R}_{k};p,q)\ge\Gamma_r,\label{eta_r_f}\\
    &\eta_c(\boldsymbol{R},\boldsymbol{R}_{k};k)\ge\Gamma_c,~ k=1,...,K ,\label{eta_c_f}\\
    & \boldsymbol{R}=\sum_{i=1}^{K+1}\boldsymbol{R}_i,\\
    & [\boldsymbol{R}]_{j,j}=P_t, ~j=1,...,T.
    \end{align}
\end{subequations}
We rewrite (\ref{eta_r_f}) and (\ref{eta_c_f}) with the following rearrangements:
\begin{subequations}
\label{objst}
    \begin{align}
        \centering
        &\Gamma_r^{-1}\boldsymbol{a}_t^H\left(\boldsymbol{R}-\sum_{k=1}^K\boldsymbol{R}_{k}\right)\boldsymbol{a}_t
    \ge\boldsymbol{\bar{a}}_{h}^{(t)H}\boldsymbol{R}\boldsymbol{\bar{a}}_{h}^{(t)}+{\sigma^2}/{\left|\bar{A}_{r}^{(t)}\right|^2},\label{etar_unequal}\\
    & (1+\Gamma_c^{-1})\boldsymbol{\hat{h}}_{k}^{(t)H}\boldsymbol{R}_k\boldsymbol{\hat{h}}_{k}^{(t)}\ge \boldsymbol{\hat{h}}_{k}^{(t)H}\boldsymbol{R}\boldsymbol{\hat{h}}_{k}^{(t)}+\sigma^2,~k=1,...,K,\label{etac_unequal}
    \end{align}
\end{subequations}
where the $\boldsymbol{\hat{h}}_{k}^{(t)}$ is the cascaded communication channel updated via (\ref{cascaded channel}). By respectively substituting rearrangements in (\ref{etar_unequal}) and (\ref{etac_unequal}) into constraints (\ref{eta_r_f}) and (\ref{eta_c_f}), problem (\ref{st5}) becomes convex and can be solved using CVX \cite{toh1999sdpt3}.

Similar to \cite{pritzker2022transmit}, the feasibility problem (\ref{st5}) has a closed-form solution
$\boldsymbol{\hat{R}}_1,...,\boldsymbol{\hat{R}}_K,\boldsymbol{\hat{R}}_{K+1}$,
satisfying:
\begin{equation}
\label{theo_rank}
    \begin{aligned}
    \centering
    {\rm rank}(\boldsymbol{\hat{R}}_i)=1, ~i=1,...,K+1.
    \end{aligned}
\end{equation}
{Obviously, for each $\Gamma_r$, if (\ref{st5}) is feasible, we have a solution that satisfies rank-one constraints in (\ref{theo_rank}). 
By bisection search on $\Gamma_r$, we denote the solution of (\ref{st5}) corresponding to the maximum value of the variable $\Gamma_r$ as the optimal solution of the original maximization problem (\ref{st2}).}
 Then, the \ac{bs}'s communication beamforming matrix $\boldsymbol{W}_c=[\boldsymbol{w}_1,...,\boldsymbol{w}_K]$ is calculated by 
\begin{equation}
\label{eq:cholesky}
    \begin{aligned}
    \centering
    \boldsymbol{\hat{w}}_k = \left(\boldsymbol{\hat{h}}_{k}^{(t)H}\boldsymbol{\hat{R}}_k \boldsymbol{\hat{h}}_{k}^{(t)}\right)^{-\frac{1}{2}}\boldsymbol{\hat{R}}_k \boldsymbol{\hat{h}}_{k}^{(t)}, ~k=1,...,K,
    \end{aligned}
\end{equation}
and the radar beamforming vector $\boldsymbol{\hat{w}}_r$ is given by
\begin{equation}
\label{eq:cholesky2}
    \begin{aligned}
    \centering
    \boldsymbol{\hat{w}}_r = \left(\boldsymbol{a}_t^H \left(\boldsymbol{\hat{R}}-\sum_{k=1}^K\boldsymbol{\hat{R}}_{k} \right)\boldsymbol{a}_t\right)^{-\frac{1}{2}}\left(\boldsymbol{\hat{R}}-\sum_{k=1}^K\boldsymbol{\hat{R}}_{k} \right) \boldsymbol{a}_t.
    \end{aligned}
\end{equation}

{In summary, to accomplish the transmitted beam design with a fixed \ac{hris} configuration, we convert the maximization problem (\ref{st2}) into a feasibility problem (\ref{st3}) of the variable $\Gamma_r$. For a fixed $\Gamma_r$, we solve the feasibility problem (\ref{st3}). By using the \ac{sdr} technique \cite{4} to relax,  all the constraints in (\ref{st5}) are linear matrix inequalities over $\boldsymbol{R}_i$ and thus can be solved via CVX \cite{toh1999sdpt3}.
Correspondingly, the optimal value of $\Gamma_r$ can be obtained by a bisection search, and the beamforming matrix $\boldsymbol{W}$ is calculated by (\ref{eq:cholesky}) and (\ref{eq:cholesky2}).}

\vspace{-0.2cm}
\subsection{Alternating Optimization strategy in HRIS-assisted  MIMO DFRC System Beamforming}
\label{sec:alt_beam}

In this section, we detail how to implement the \ac{ao} of the \ac{bs} waveforms and \ac{hris} beam patterns, as well as the mechanism for updating system parameters throughout this \ac{ao} process in Algorithm \ref{al2}. 
\begin{algorithm}
    \caption{\ac{hris}-assisted \ac{mimo} \ac{dfrc} system alternating optimization}
    \renewcommand{\algorithmicrequire}{\textbf{Input:}}
    \renewcommand{\algorithmicensure}{\textbf{Output:}}
    \label{al2}
    \begin{algorithmic}[1]
        \REQUIRE \leavevmode \\
        User's \ac{sinr} threshold $\Gamma_c$;\\
        Initialization: $\boldsymbol{\beta}^0=\boldsymbol{0}_{N\times 1}$;\\
        Set the objective function $f(\boldsymbol{\beta})$ based on (\ref{objlag}).
        \REPEAT
        \STATE Compute the optimal $\boldsymbol{\beta}$ by solving the non-convex optimization problem (\ref{obj}) according to Algorithm~\ref{al1_agd} and Algorithm~\ref{al1}.
        \STATE {Calculate the reflected vector $\boldsymbol{\hat{a}}_h$ and the received scalar $A_r$ via (\ref{cascaded reflect}).}
        \STATE Compute the covariance matrix $\boldsymbol{R}$ and its sub-covariance matrices $\boldsymbol{R}_i$ of the \ac{bs} by solving (\ref{st2}) with bisection search and \ac{sdr} techniques.
        \STATE Compute $\boldsymbol{w}_i$ via (\ref{eq:cholesky}) and (\ref{eq:cholesky2}).
        \STATE Update the matrices $\boldsymbol{C}_1$, $\boldsymbol{C}_2$, $\boldsymbol{C}_3$ via (\ref{C_1_2_3}).
        \UNTIL $ \left|f\left(\boldsymbol{\beta}^{(t)}\right)-f\left(\boldsymbol{\beta}^{(t-1)}\right)\right| < \epsilon$
        \STATE Select the total beamforming matrix $\boldsymbol{W}=[\boldsymbol{w}_1,..., \boldsymbol{w}_{r}]$
        \ENSURE \leavevmode \\
        The \ac{hris} configuration vector $\boldsymbol{\beta}$;\\
        The \ac{bs} beamforming matrix $\boldsymbol{W}$.
    \end{algorithmic}
\end{algorithm}


First, as noted in solving (\ref{obj}), {the configuration of the \ac{hris} is determined utilizing the \ac{fgs}-\ac{agd} approach. Second, we use this configuration to compute the reflected vector $\boldsymbol{\hat{a}}_h$ and received scalar $A_r$ according to (\ref{cascaded reflect})}, which are used to address the joint transmitted beamforming design (\ref{st2}). Then, {we use the bisection search and \ac{sdr} techniques} to obtain the covariance matrix $\boldsymbol{R}$ by tackling the optimization problem (\ref{st2}).

Note that the propagation matrices $\boldsymbol{C}_1$, $\boldsymbol{C}_2$, $\boldsymbol{C}_3$ will change with the \ac{bs} beamforming matrix $\boldsymbol{W}=[\boldsymbol{W}_c,\boldsymbol{w}_{r}]$, so that we need to update these system parameters during iterations. Combining with the definitions of the matrices $\boldsymbol{C}_1$, $\boldsymbol{C}_2$, and $\boldsymbol{C}_3$, we can update these propagation matrices according to (\ref{C_1_2_3}).

As we complete updating the \ac{hris}-assisted \ac{mimo} \ac{dfrc} system parameters, we need to optimize the \ac{hris} configuration as well as the joint transmitted beam design of the \ac{bs} continually. After several iterations, the Algorithm~\ref{al2} stops when $ \left|f\left(\boldsymbol{\beta}^{(t)}\right)-f\left(\boldsymbol{\beta}^{(t-1)}\right)\right|<\epsilon$.

\subsection{Complexity and Convergence Analysis}
    \subsubsection{Computational Complexity Analysis} 
    \begin{itemize}
        \item Optimization of \acp{bs}'s beamforming matrices $\boldsymbol{W}_c$, $\boldsymbol{w}_r$ with fixed \ac{hris} configuration: The \ac{sdp} problem in (\ref{st5}) is solved by the \ac{sdr} approach during a bisection search. Specifically, given the potential values of $\Gamma_r$, the number of execution steps of the bisection search $\log_2 {(\frac{\Gamma_{r,max}-\Gamma_{r,min}}{\epsilon_1})}$. Similarly, given the solution accuracy $\epsilon_2$, the complexity of using the \ac{sdr} to solve (\ref{st5}) is $\mathcal{O}\left(T^{3.5}\log(1/\epsilon_2)\right)$ \cite{luo2010semidefinite}. Therefore, the computational complexity of \acp{bs}'s beamforming algorithm is $\log_2 {(\frac{\Gamma_{r,max}-\Gamma_{r,min}}{\epsilon_1})}\times \mathcal{O}\left(T^{3.5}\log(1/\epsilon_2)\right)$.

        \item Optimization of \ac{hris} configuration with fixed \acp{bs} beamforming matrices: Since we suggest the fast grid search assisted auto gradient descent algorithm to solve the problem in (\ref{objlag}), the complexity of this algorithm is $M_{max} Z_{max}\times \mathcal{O}\left( M N^2\right )$, where $ M_{max} Z_{max}$ is the execution times of the fast grid search, and $\mathcal{O}\left( M N^2\right )$ is the complexity of using auto gradient descent for a fixed maximum step $M$.
    \end{itemize}
    
    As a result, the computational complexity of the proposed alternating algorithm for each iteration is $\log_2 {(\frac{\Gamma_{r,max}-\Gamma_{r,min}}{\epsilon_1})}\times \mathcal{O}\left(T^{3.5}\log(1/\epsilon_2)\right)+M_{max} Z_{max}\times \mathcal{O}\left( M N^2\right )$.    
    \subsubsection{Convergence Analysis}
    Algorithm~3 explains an alternating process, it is necessary to analyze the convergence of each sub-algorithm and the whole \ac{ao}.
    \begin{itemize}
        \item The ascent property of the \ac{fgs}-\ac{agd}: Since the \ac{fgs} is a grid search-based algorithm, it obviously will converge. Next, we prove the convergence of the \ac{agd}. The objective function $f(\boldsymbol{\beta})$ is L-smooth, then \ac{agd} with the step size $\Delta_i=1/L$ obeys
        \begin{equation}
        \label{gd}
            \begin{aligned}
            \centering
                f\left(\boldsymbol{\beta}^{i+1}\right) \leq f\left(\boldsymbol{\beta}^i\right)-\frac{1}{2 L}\left\|\nabla f\left(\boldsymbol{\beta}^i\right)\right\|_2^2.
            \end{aligned}
        \end{equation}
        When the step size $\Delta_i$ is sufficiently small, \ac{agd} can guarantee a decrease in the value of the objective function $f(\boldsymbol{\beta})$. After the maximum step $M$, yielding
        \begin{equation}
        \label{M_ite}
            \begin{aligned}
            \centering
            f\left(\boldsymbol{\beta}^{M}\right) &\leq f\left(\boldsymbol{\beta}^0\right)-\frac{1}{2 L}\sum_{i=0}^{M-1}\left\|\nabla f\left(\boldsymbol{\beta}^i\right)\right\|_2^2\\
            &\leq f\left(\boldsymbol{\beta}^0\right)-\frac{M}{2 L}\left\|\nabla f\left(\boldsymbol{\beta}^i\right)_{min}\right\|_2^2.
            \end{aligned}
        \end{equation}
        Therefore, the objective function during two iterations of the \ac{ao} holds that
        \begin{equation}
        \label{t_ite}
            \begin{aligned}
                \centering
                f\left(\boldsymbol{W}^{t},\boldsymbol{\beta}^{t+1}\right) \leq f\left(\boldsymbol{W}^{t},\boldsymbol{\beta}^{t}\right)-\frac{M}{2 L}\left\|\nabla f\left(\boldsymbol{W}^{t},\boldsymbol{\beta}\right)_{min}\right\|_2^2.
            \end{aligned}
        \end{equation}
        The inequality in (\ref{t_ite}) verifies that the objective value in the $t+1$-th iteration is lower than the $t$-th iteration.

        \item The ascent property of the \ac{sdr} with bisection search: As the goal of (\ref{st2}) is to maximize the radar \ac{sinr} by optimizing the beamforming matrix of the \acp{bs} with the fixed $\boldsymbol{\beta}$, we have $\eta_r(\boldsymbol{W}^{t+1},\boldsymbol{\beta}^{t+1})\geq \eta_r(\boldsymbol{W}^{t},\boldsymbol{\beta}^{t+1})$. However, the objective value is defined by the negative radar \ac{sinr}, yielding
        \begin{equation}
        \label{wbeta}
            \begin{aligned}
                \centering
                f\left(\boldsymbol{W}^{t+1},\boldsymbol{\beta}^{t+1}\right) \leq f\left(\boldsymbol{W}^{t},\boldsymbol{\beta}^{t+1}\right).
            \end{aligned}
        \end{equation}
        
        \item The convergence of the proposed \ac{ao}:
        In addition to the ascent property of both sub-algorithms, the convergence of the proposed \ac{ao} also relies on the boundary. Combining (\ref{t_ite}) with (\ref{wbeta}), we have
        \begin{equation}
        \label{twbeta}
            \begin{aligned}
                \centering
                f\left(\boldsymbol{W}^{t+1},\boldsymbol{\beta}^{t+1}\right) &\leq f\left(\boldsymbol{W}^{t},\boldsymbol{\beta}^{t}\right)\\
                &-\frac{M}{2 L}\left\|\nabla f\left(\boldsymbol{W}^{t},\boldsymbol{\beta}\right)_{min}\right\|_2^2.
            \end{aligned}
        \end{equation}
        Due to the per-antenna power constraint of the \acp{bs} in (\ref{h_poewr}) and the \ac{hris} power limit in (\ref{b_power}), the radar \ac{sinr} $\eta_r(\boldsymbol{W},\boldsymbol{\beta})$ is upper-bounded so that the objective value $f$ is lower-bounded. Consequently, the Algorithm~3 is convergent and holds:
        \begin{equation}
        \label{converge}
            \begin{aligned}
                \centering
                \left|f\left(\boldsymbol{W}^{t+1},\boldsymbol{\beta}^{t+1}\right)-f\left(\boldsymbol{W}^{t},\boldsymbol{\beta}^{t}\right)\right|\leq \epsilon.
            \end{aligned}
        \end{equation}
    \end{itemize}

	\vspace{-0.2cm}
	\section{Numerical Evaluations}
	\label{sec:Sims}
	
In this section, we provide numerical results to evaluate the performance of our proposed joint beamforming design algorithm for \ac{hris}-assisted \ac{mimo} \ac{dfrc} systems. Specially, we first describe the simulation setup of the \ac{hris}-assisted \ac{mimo} \ac{dfrc} system in section \ref{subsec:setup} and then present some numerical results to analyze the improvements in radar and communication performance compared with the \ac{bs}-only system and \ac{bs}-\ac{ris} system in section~\ref{subsec:converg}. 

\vspace{-0.2cm}
\subsection{Simulation Setup}
\label{subsec:setup}

The proposed \ac{hris}-assisted \ac{mimo} \ac{dfrc} configuration is illustrated in Table. \ref{tab:hrisplat}, where the wavelength $\lambda=0.1$~m. The total number of elements in \ac{hris} is $N=16$, which is a square planar array deployed on the 3D Cartesian coordinate's YOZ plane. The radar signal $s(n)$ is generated by random phase coding, and the communication symbols are generated by 16-quadrature
amplitude modulation (QAM) modulation. To perform the \ac{ao} approach for the formulated optimization problem, we defined the hyper-parameters in the \ac{hris} optimization as~$\lambda_1=4$, $\lambda_2=0.8$, and give the range of potential values of $\Gamma_r$ as~ $\Gamma_{r,max}=38$~dB, $\Gamma_{r,min}=22$~dB.
    
\begin{table}[htbp]
    \centering
    \caption{The \ac{dfrc} platform simulation configuration}
    \begin{tabular}{c|c}
    \hline \hline
        Parameters & Value \\
        \hline
        Center frequency & $3$~GHz \\
        The inter-element spacing in \ac{hris} & $\lambda$ \\
        The inter-element spacing in \ac{bs} & $\lambda/2$ \\
        Position of the \ac{hris} center & $(0, 100\lambda, 30\lambda)$ \\
        Position of the \ac{bs} center& $(0, 0, 300\lambda)$ \\
        Position of the user & $(75\lambda, 100\lambda, 0)$ \\
        Position of the target & $(0,0,0)$\\
        $\Gamma_c$ & 5 dB \\
        $N$ & 16\\
        $T$ & 8\\
        \hline \hline
    \end{tabular}
    \label{tab:hrisplat}
\end{table}


\vspace{-0.2cm}
\subsection{Performance Analysis}
\label{subsec:converg}

\begin{figure}[htbp]
	\centering  
	\subfigtopskip=2pt 
	\subfigbottomskip=2pt 
	\subfigcapskip=-5pt 
	\subfigure[Convergence of the proposed Algorithm~\ref{al2}.]{
		\label{convergence_aia}
		\includegraphics[width=0.45\linewidth]{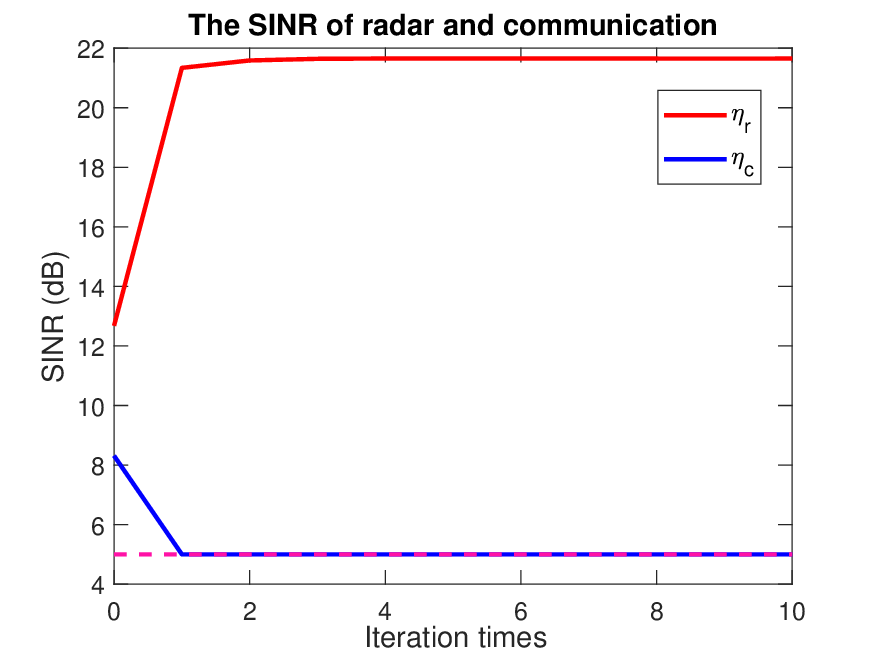}}
	\subfigure[Convergence of the \ac{fgs}-\ac{agd} algorithm.]{
		\label{fig:agdsinr}
		\includegraphics[width=0.45\linewidth]{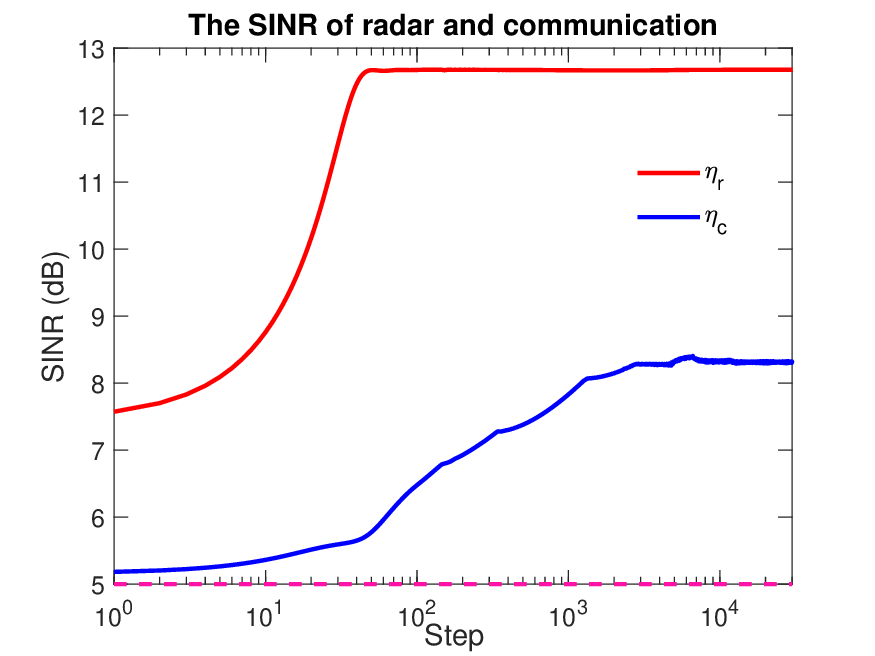}}
	\caption{Convergence performance of the \ac{ao} approach in Algorithm~\ref{al2} and the \ac{fgs}-\ac{agd} algorithm. The red line is the radar's \ac{sinr}, the blue line is the communication user's \ac{sinr} and the pink dash is the threshold of communication.}
	\label{convergence}
\end{figure}

First, we study the convergence performances of both the \ac{ao} approach in Algorithm~\ref{al2} and the \ac{fgs}-\ac{agd} algorithm, where the user is deployed at $(75\lambda, 100\lambda, 0)$ and the radar target is put at 3D Cartesian coordinate center $(0, 0, 0)$. In particular, by applying the \ac{fgs}-\ac{agd} for \ac{hris} configuration optimization and the \ac{sdr} technique for \ac{bs} beamforming design, the convergence performance is shown in Fig. \ref{convergence_aia}, when the per-antenna power of the \ac{bs} is $P_t = 0$~dB.
It indicates the proposed algorithm achieves convergence very fast in less than four iterations. 
In addition, the convergence of the proposed \ac{fgs}-\ac{agd} method is shown in Fig. \ref{fig:agdsinr}. With the assistance of the initialization operation in Algorithm~\ref{al1}, Algorithm~\ref{al1_agd} converges to a fixed value within $1000$ iterations, which shows the efficiency and stability of the \ac{fgs}-\ac{agd} algorithm.

\begin{figure}[htbp]
	\centering  
	\vspace{-0.35cm} 
	\subfigtopskip=2pt 
	\subfigbottomskip=2pt 
	\subfigcapskip=-5pt 
	\subfigure[Consistent.]{
		\label{fig:same_value}
		\includegraphics[width=0.45\linewidth]{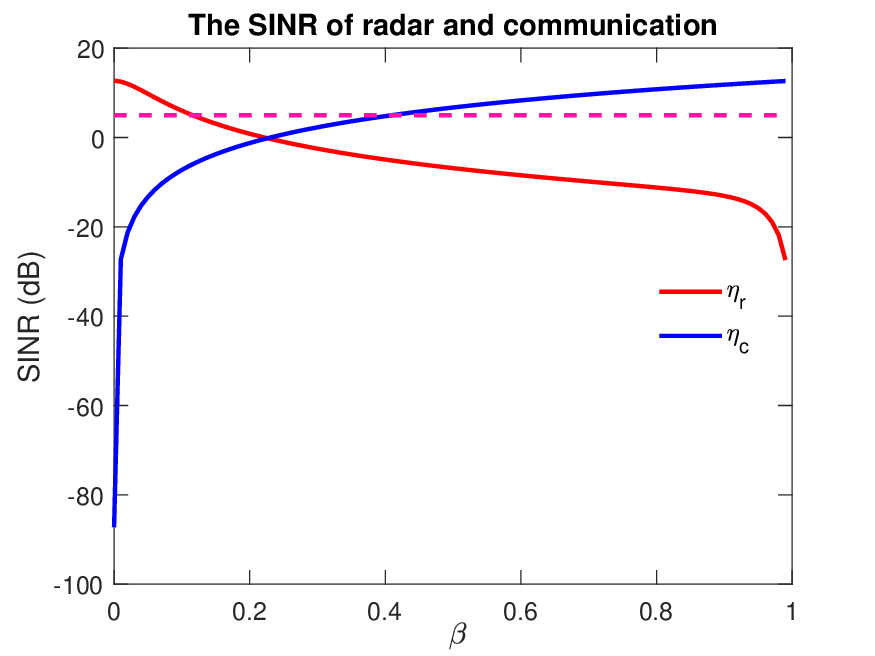}}
	\subfigure[Random.]{
		\label{fig:rand_value}
		\includegraphics[width=0.45\linewidth]{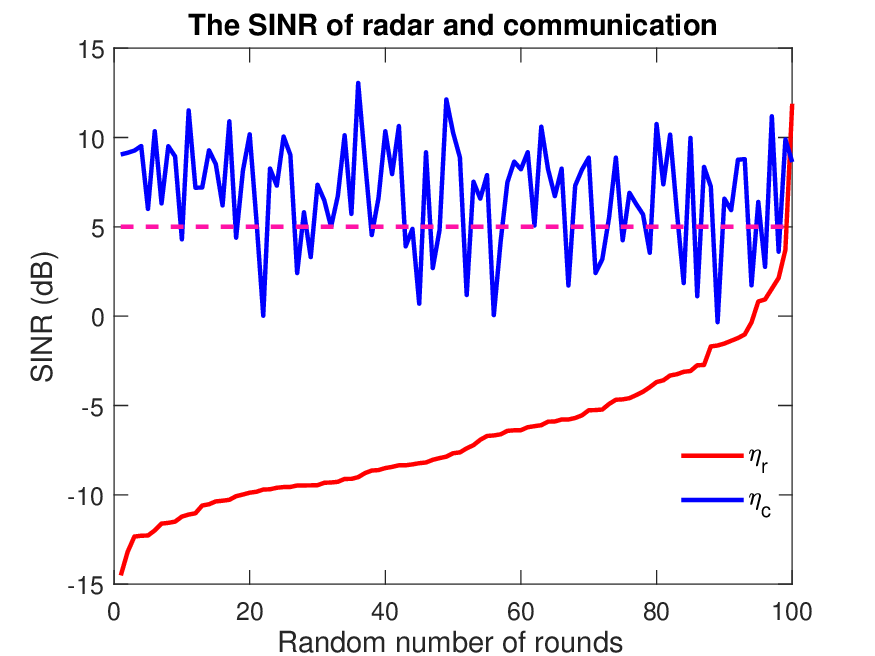}}
	\caption{The \ac{sinr} of radar and communication. The red line represents the \ac{sinr} of radar, the blue line is the \ac{sinr} of the user, and the pink dash shows the threshold of the communication (a) All elements of the \ac{hris} have the same value. (b) The $\boldsymbol{\beta}$ is a random sequence.}
	\label{baseline}
\end{figure}

To evaluate the effect of the \ac{hris} configuration on the \ac{dfrc}'s performance, we explore the radar’s and communication’s \ac{sinr} behaviors via different amplitude distributions on the \ac{hris}. In Fig. \ref{fig:same_value}, all the elements in the \ac{hris} are set to have the same value, and the value of $\boldsymbol{\beta}$ varies between $0$ and $1$.  
It shows that the \ac{sinr} of the radar will degrade when the \ac{sinr} of the user tries to satisfy our constraints. 
Since the communication signal reflection and radar echo reception are determined by the operation of the \ac{hris} in splitting the power of each element attached. There will be a critical trade-off between the performance of radar and communication. Fig.~\ref{fig:same_value} verifies the essential trade-off between the \ac{sinr} performance of the radar and communication.
In Fig. \ref{fig:rand_value}, we randomize the value of $\boldsymbol{\beta}$, and get good performance in both radar and communication at some points. This demonstrates that the \ac{sinr} performance of both radar and communication will be enhanced by \ac{hris} configuration design.

We numerically evaluate the performance gain of our designed \ac{hris}-assisted \ac{mimo} \ac{dfrc} system with the \ac{bs}-only and random-configured \ac{hris} systems and compare the proposed \ac{fgs}-\ac{agd} algorithm with a conventional optimization approach, such as \ac{ga}. 
\begin{table*}[htbp]
    \centering
    \caption{The variables and configuration of each competitor}
    \begin{tabular}{c|ccc|c}
\hline
BS-only & \multicolumn{3}{c|}{HRIS-assisted}                                                                                                               & Profile                 \\ \hline
SDR     & \multicolumn{1}{c|}{GA}                   & \multicolumn{1}{c|}{AGD-FGS}                & Arbitrary selection& Algorithm               \\
5 dB    & \multicolumn{1}{c|}{5 dB}                 & \multicolumn{1}{c|}{5 dB}                   & 5 dB   & Communication Threshold \\
8       & \multicolumn{1}{c|}{8}                    & \multicolumn{1}{c|}{8}                      & 8      & The number of BS        \\
-       & \multicolumn{1}{c|}{16}                   & \multicolumn{1}{c|}{16}                     & 16     & The number of HRIS      \\
$\boldsymbol{W}$       & \multicolumn{1}{c|}{$\boldsymbol{W}$, $\boldsymbol{\beta}$} & \multicolumn{1}{c|}{$\boldsymbol{W}$, $\boldsymbol{\beta}$} & $\boldsymbol{W}$      & Optimized variables     \\
$P_t$  & \multicolumn{1}{c|}{$P_t$} &
\multicolumn{1}{c|}{$P_t$} & $P_t$     & Transmit power     \\ \hline
    \end{tabular}
    \label{tab:competitors}
\end{table*}
Specifically, the \ac{bs}-only system performs both radar sensing and communication using the same hardware, and its \ac{bs} has the same architecture as the proposed system. 
The random-configured \ac{hris} system utilizes the \ac{hris} to reflect or receive the arriving signals to perform communication with mobile users and radar sensing. 
For the algorithmic comparisons, we compare the proposed \ac{fgs}-\ac{agd} with the traditional \ac{ga}, and arbitrary configuration. Note that we optimize the beamforming matrix $\boldsymbol{W}$ for the \acp{bs}-only system and random-configured \ac{hris}-assisted system. As shown in Table.~\ref{tab:competitors}, we fairly compare those four configurations in terms of the communication threshold, transmit power, the number of elements at the \ac{bs} and \ac{hris}, and the optimized variables.

\begin{figure}
    \centering
    \includegraphics[width=0.95\linewidth]{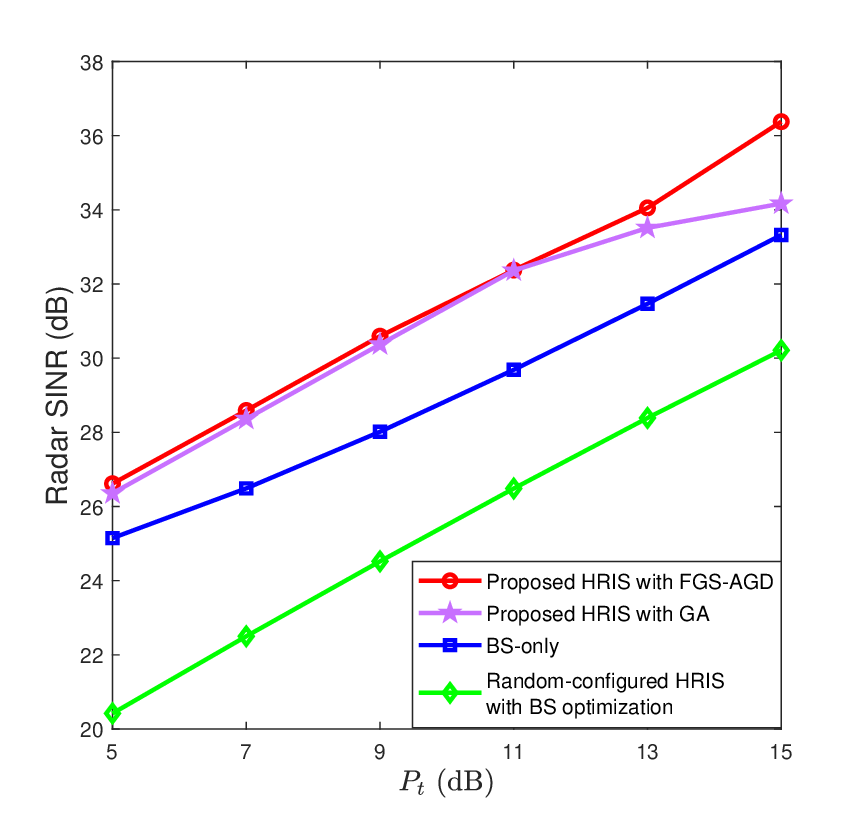}
    \caption{Comparison of the radar \ac{sinr} of three types of \ac{dfrc} systems under different per-antenna power supplies.}
    \label{fig:com_dfrc_sinr}
\end{figure}
In Fig. \ref{fig:com_dfrc_sinr}, compared with the \ac{bs}-only and the random-configured \ac{hris} systems, the proposed \ac{hris}-assisted \ac{mimo} \ac{dfrc} system obtains the best radar performance with the same transmit power. 
This demonstrates that the optimized \ac{hris} configuration improved the performance of radar sensing while assuring the communication requirement. Moreover, the random-configured \ac{hris}-assisted system, designed by \acp{bs} beamforming rather than alternately optimizing parameters of \ac{hris} and \ac{bs}, does not achieve its best performance. The reason for this is that the \ac{hris} can adjust the EM environment by configuring its power splitting factor to facilitate radar sensing and communication.
Then we analyze the effectiveness and robustness of the proposed algorithm. Fig.~\ref{fig:com_dfrc_sinr} shows that using the \ac{fgs}-\ac{agd} algorithm can achieve better radar performance than using conventional \ac{ga} for the joint problem of the \ac{bs}'s beamforming design and \ac{hris} configuration, which verifies the effectiveness and robustness of the \ac{fgs}-\ac{agd}.

In order to comprehend the benefit of the \ac{hris} configuration design and the \ac{bs} transmitted beam design, we provide 2D beampattern results of the \ac{hris} by employing the proposed \ac{fgs}-\ac{agd} to demonstrate the power allocation impact of the radar waveform and communication waveform of the \ac{bs}. There is a concise explanation for beampattern generations of the \ac{hris} and \acp{bs}.
    \begin{itemize}
        \item Beampattern of the \ac{hris}:
        Given the configured $\boldsymbol{\beta}$, the beampattern of the \ac{hris} is generated by
    \begin{equation}
    \label{AF_HRIS}
        \begin{aligned}
            \centering
            \mathrm{AF}(\theta, \phi)=&\sum_{m=0}^{N_x-1} \sum_{n=0}^{N_y-1} w(m, n) \\
            &\exp \left(j \frac{2 \pi}{\lambda} m d x \sin \theta \cos \varphi\right) \\
            &\exp \left(j \frac{2 \pi}{\lambda} n d y \sin \theta \sin \varphi\right),
        \end{aligned}
    \end{equation}
    where $w(m,n)=\beta_l$, for $m=l\bmod N_x,~n=\lfloor l/N_x \rfloor +1 $, is the inspiring power corresponding to the value of power splitting factor $\boldsymbol{\beta}$ with the number of elements being $N_xN_y=N$.

        \item Beampattern of the \acp{bs}:
        Given the precoders of the \acp{bs}, we can define the beampatterns related to precoders of radar detection and communication as
    \begin{equation}
        \begin{aligned}
            \centering
            \mathrm{B}_r(\theta)=\frac{\left \Vert \boldsymbol{a}^H(\theta)\boldsymbol{W}_r\right \Vert_2^2}{\mathrm{tr}(\boldsymbol{W}_r\boldsymbol{W}_r^H)},
        \end{aligned}
    \end{equation}
    \begin{equation}
        \begin{aligned}
            \centering
            \mathrm{B}_c(\theta)=\frac{\left \Vert \boldsymbol{a}^H(\theta)\boldsymbol{W}_c\right \Vert_2^2}{\mathrm{tr}(\boldsymbol{W}_c\boldsymbol{W}_c^H)},
        \end{aligned}
    \end{equation}
    where $\boldsymbol{a}(\theta)$ is the transmit steering vector in direction $\theta$.
        
    \end{itemize}

\begin{figure*}[ht]
	\centering  
	\vspace{-0.35cm} 
	\subfigtopskip=2pt 
	\subfigbottomskip=2pt 
	\subfigcapskip=-5pt 
	\subfigure[Beampattern on the YOZ plane.]{
		\label{fig:radar_gua}
		\includegraphics[width=0.95\linewidth]{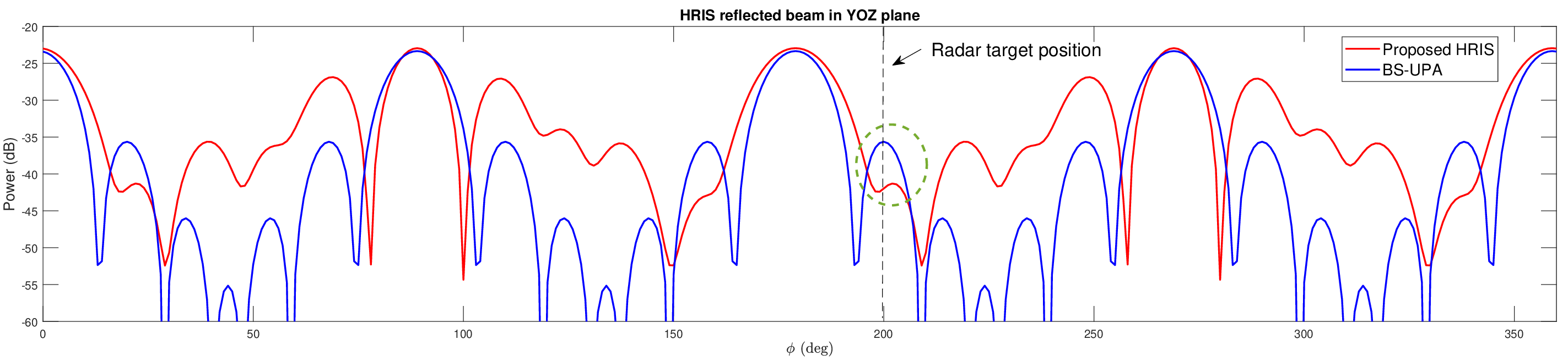}}
		
	\subfigure[Beampattern on the $\phi=90^\circ$ plane.]{
		\label{fig:user_gua}
		\includegraphics[width=0.95\linewidth]{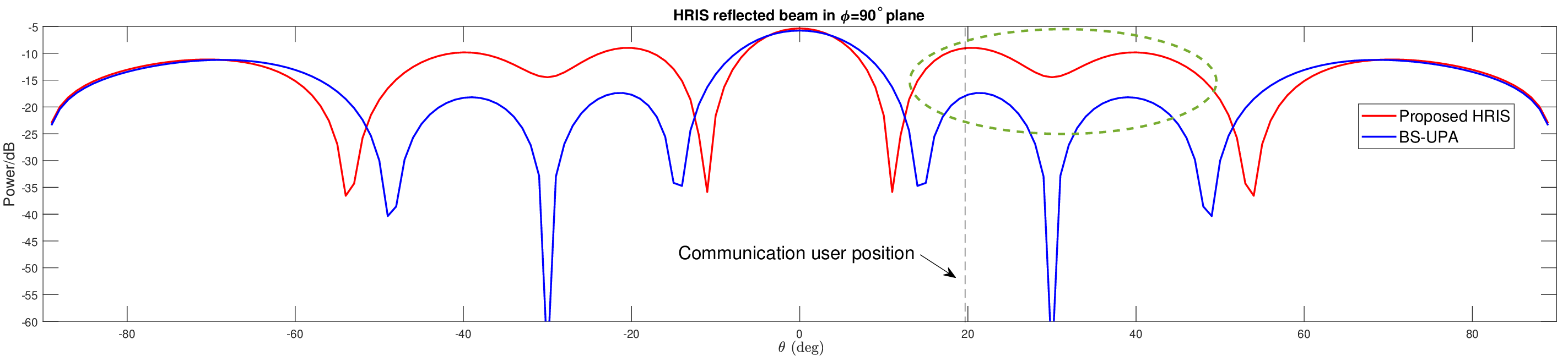}}
	\caption{The comparison of the benchmark and that designed by \ac{fgs}-\ac{agd}.}
	\label{jrc_gua}
\end{figure*}
To understand what we get from this \ac{hris} configuration design, we analyze the visual results of the \ac{hris} beampattern and investigate the effects of the radar target's location and the communication user's position on \ac{hris} configuration design. 
In particular, we explore the correspondence between the beampattern direction modified by the \ac{hris} configuration and the \ac{dfrc} performance. To this aim, the beams of the \ac{hris} radiation pattern in the YOZ plane and the $\phi=90^\circ$ plane are illustrated in Fig. \ref{fig:radar_gua} and Fig. \ref{fig:user_gua}, respectively. 
Assuming that rotation angle $\phi$ is formed by rotation along the $y$ axis to the $z$ axis, and $\theta$ is the look-down angle of the \ac{hris}, Fig. \ref{jrc_gua} compares the beam patterns of the \ac{ris} with that the proposed \ac{hris} designed by \ac{fgs}-\ac{agd}. We observe the effect of the radar target's location on the \ac{hris} beam design on the YOZ plane and calculate that the rotation angle $\phi$ of the target relative to the \ac{hris} is around $198^\circ$. In Fig. \ref{fig:radar_gua}, the beam designed by \ac{fgs}-\ac{agd} reduces the side-lobe power level at $\phi\approx 200^\circ$, and results in the main-lobe widening, which means that the optimized \ac{hris} configuration can reduce the interference of the \ac{hris} reflected beam to the detection zone. Then, we study the effect of the communication user's position on the \ac{hris} reflected beam design on the $\phi=90^\circ$ plane and calculate that the look-down angle of the users is around $20^\circ$. We find a similar outcome in Fig. \ref{fig:user_gua}. The beam created by \ac{fgs}-\ac{agd} increases the side-lobe power level at the look-down angle of $\theta\approx 20^\circ$, and at the same time reduces the main lobe, which indicates that the optimized \ac{hris} configuration increases the effective power of communication user by increasing the corresponding side-lobe power level. Hence, we infer that the \ac{hris} adjusts the side-lobe of the beampattern by modifying the amplitude distribution of the surface components, thus increasing the quality of radar detection while maintaining the quality of communication.
\begin{figure}[htp]
	\centering  
	\vspace{-0.35cm} 
	\subfigtopskip=2pt 
	\subfigbottomskip=2pt 
	\subfigcapskip=-5pt 
	\subfigure[\ac{ga}.]{
		\label{fig:gaafbs}
		\includegraphics[width=0.45\linewidth]{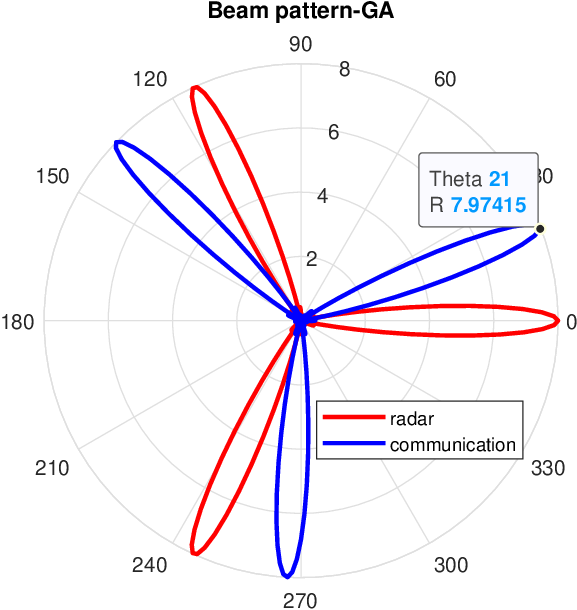}}
	\subfigure[\ac{fgs}-\ac{agd}.]{
		\label{fig:agdafbs}
		\includegraphics[width=0.45\linewidth]{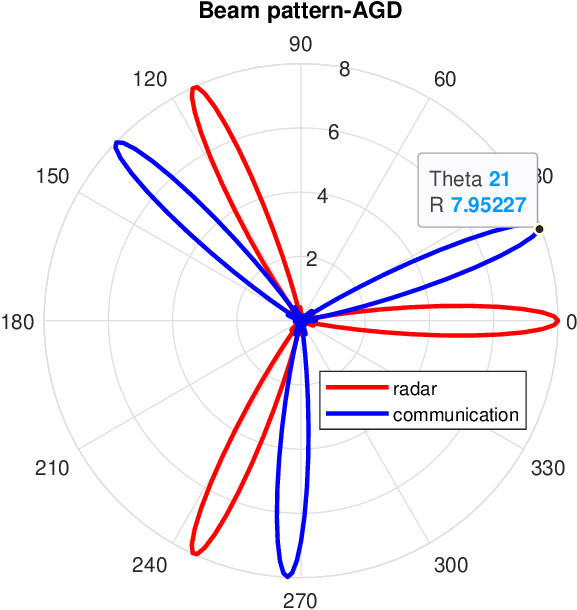}}
	\caption{Optimized joint transmitted beampattern. (a) shows the communication and radar waveforms of the \ac{bs} by using the \ac{ga} to obtain the amplitude distribution of \ac{hris}, and (b) shows the joint transmitted waveforms by using the \ac{fgs}-\ac{agd} to design the \ac{hris} beampattern.}
	\label{transbeam}
\end{figure}
\begin{figure}[htp]
    \centering
    \includegraphics[width=0.95\linewidth]{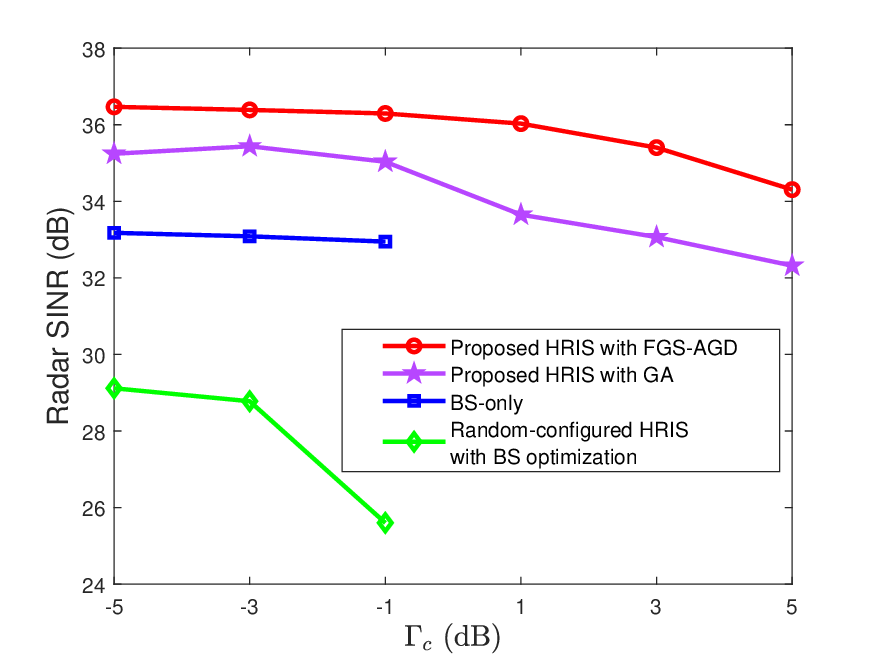}
    \caption{Comparison of the \ac{sinr} of the radar under different communication thresholds.}
    \label{fig:maxsinr_com}
\end{figure}

Next, in the joint transmitted beam design, we employ the optimized \ac{hris} configuration.
Since we provide the \ac{fgs}-\ac{agd} approach and the \ac{ga} to accomplish the \ac{hris} configuration optimization, there are two types of amplitude distributions of the \ac{hris}. Based on each solution of \ac{hris} beam design, we are able to calculate an optimal solution of the joint transmitted beam design.
In Fig. \ref{transbeam}, we compare the optimization results of the joint transmitted beamforming for two different \ac{hris} amplitude distributions. By geometric calculation, the center of the \ac{hris} is located at a position of about $20^\circ$ of the look-down angle on the \ac{bs}, and this angle corresponds precisely with the direction of the communication beam we designed. Meanwhile, the look-down angle of the radar beam is around $0^\circ$ and points to the center of the detection zone, where our deployed radar target is placed. The fact shows that the joint beam design of the \ac{bs} realizes the spatial diversity of the radar and communication waveforms and that the two beam patterns will not conflict in space.

To further evaluate the impact of spatial diversity on the performance gain of the proposed \ac{dfrc} system, we explored the radar performance in a multi-user scenario. Specifically, we evaluate the radar \ac{sinr} variation under different communication thresholds $\Gamma_c$ in Fig. \ref{fig:maxsinr_com}, considering deploying another communication user at $(150\lambda, 100\lambda, 0)$. In this scenario, the transmit power of per-antenna on the \ac{bs} is set as $P_t=15$~dB and the $\Gamma_c$ varies from $-5$~dB to $5$~dB with an increasing step set at $2$~dB. 
From Fig.~\ref{fig:maxsinr_com}, we can see that with the communication threshold $\Gamma_c$ rising the radar’s \ac{sinr} decreasing, which demonstrates a clear trade-off between radar sensing and communication.
In addition, we note that the solutions of the \ac{bs}-based and random-configured \ac{hris}-based \ac{dfrc} systems disappeared when the communication threshold is over $1$~dB. This phenomenon is reasonable because the second user is sheltered by the first user in the view of \ac{bs}. However, our proposed system modifies the \ac{em} environment and adjusts the beam to illuminate the second user. Thus, from numerical results, our proposed \ac{hris}-assisted \ac{dfrc} system achieves the highest radar \ac{sinr} than others’ presented \ac{dfrc} systems. Specifically, the radar's \ac{sinr} improves $3$~dB than the \ac{bs}-only system and $6$~dB over the random-configured \ac{hris}-based system under the same communication threshold and same transmitted power.

Finally, we evaluate the impact of the number of elements in the \ac{hris} on the radar \ac{sinr} and preliminary analyze the influence on the radar performance of the cluttered environment and imperfect \ac{csi} of communication users.

\begin{figure}[htp]
	\centering  
	\vspace{-0.35cm} 
	\subfigtopskip=2pt 
	\subfigbottomskip=2pt 
	\subfigcapskip=-5pt 
	\subfigure{
		\label{fig:num_RIS}
		\includegraphics[width=0.45\linewidth]{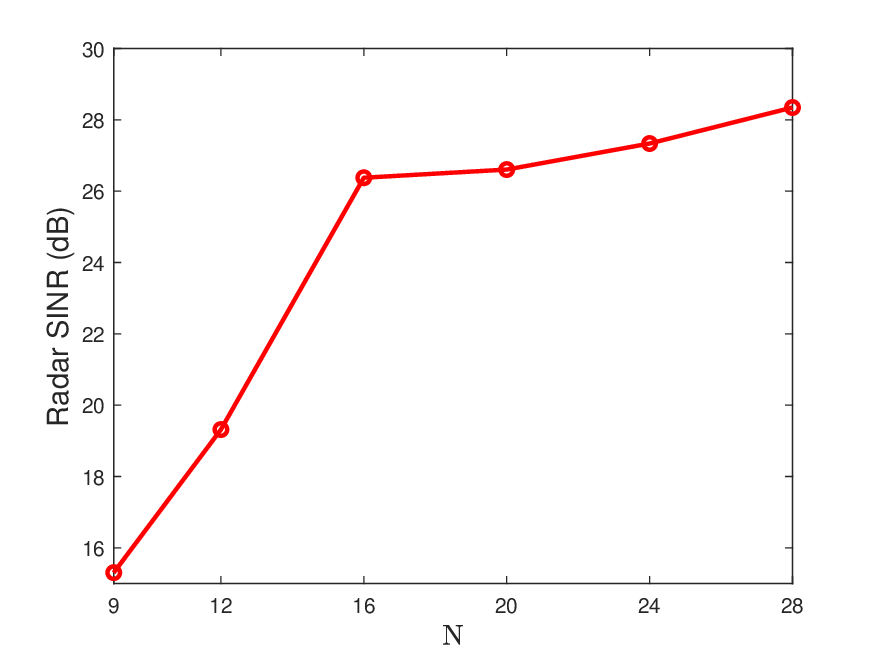}}
	\subfigure{
		\label{fig:clutter}
		\includegraphics[width=0.45\linewidth]{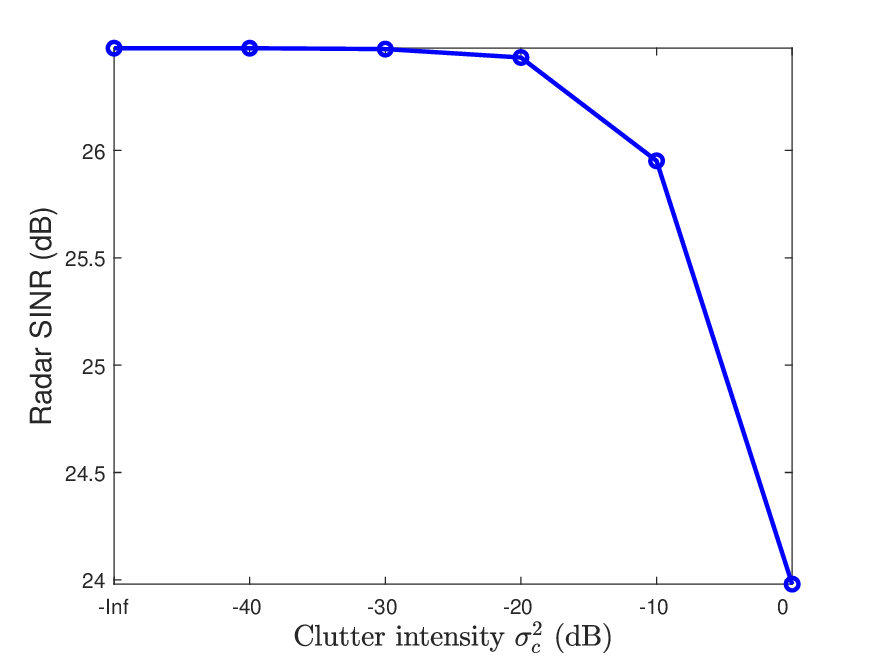}}
	\caption{Radar sensing performance. (a) Radar \ac{sinr} versus the number of the \ac{hris} elements. (b) \ac{sinr} for radar versus clutter intensity $\sigma_c^2$.}
	\label{radarper}
\end{figure}

We present the radar \ac{sinr} results for different \ac{hris} configurations under the same communication threshold $\Gamma_c=5$~dB with the transmit power of the \ac{bs} being $P_t=5$~dB. Specifically, the number of the \ac{hris} elements varies from $12$ to $28$, where the horizontal number of \ac{hris} is fixed at $4$ and the vertical number of the \ac{hris} gradually increases. We also provide the radar performance of \ac{hris} configuration with a square architecture, represented as $N=9$ elements. Fig.~\ref{fig:num_RIS} effectively demonstrates that the radar \ac{sinr} will increase as the \ac{hris}'s elements since the \ac{hris} is capable of managing more power for radar echo reception. However, it's worth noting that the increasing number of elements on the \ac{hris} also leads to the high complexity of \ac{hris} configuration optimization.

    For a cluttered environment, we consider the target scene where a single target is located in the detecting zone center, and the spatial clutter is deployed at the left side of the radar target. In particular, we consider the relative clutter intensity $\sigma_c^2$, defined as
\begin{equation}
    \begin{aligned}
        \centering
        \sigma_c^2 = P_u/P_o
    \end{aligned}
\end{equation}
where $P_u$ and $P_o$ are the original clutter power and the radar target power, respectively.

Fig.~\ref{fig:clutter} gives the radar \ac{sinr} under different clutter intensity $\sigma_c^2$. Here, $\sigma_c^2=-\infty$~dB~($0$) means that there is no clutter, and $\sigma_c^2=0$~dB means that the clutter is as strong as the radar target. It can be observed that the maximum radar \ac{sinr} may degrade when $\sigma_c^2$ increases. Comparing the results for $\sigma_c^2=-\infty$~dB~($0$) and $\sigma_c^2=0$~dB, the spatial clutter only has a minor effect on the radar sensing performance even if the relative intensity is comparable to the target. In the case of a more complex and realistic cluttered environment, the adaptation of radar beamforming to the environment is a significant topic. However, our work is only conceptual research at this stage, and in the future, we will work on the \ac{hris} \ac{dfrc} system with adaptive radar beamforming to overcome the cluttered environment.

To investigate the impact of the imperfect \ac{csi} on radar and communication \ac{sinr}, we provide numerical results of the radar \ac{sinr} and communication \ac{sinr} with imperfect \ac{csi} under different angle bias $\Delta_{\theta}$.  
\begin{figure}[htbp]
    \centering
    \includegraphics[width=.95\linewidth]{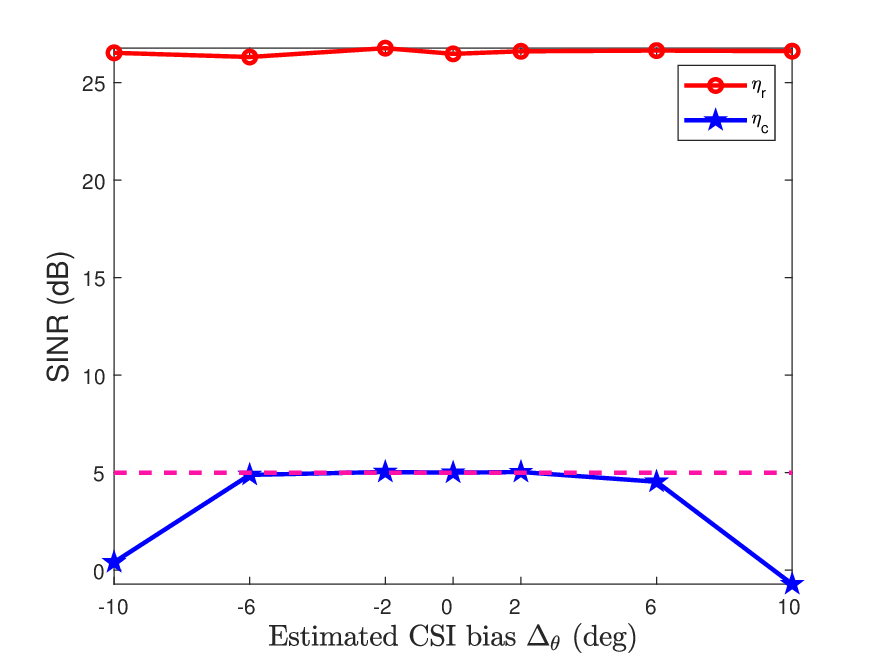}
    \caption{\acp{sinr} for radar and communication versus imperfect \ac{csi} bias.}
    \label{fig:imperfect_CSI}
\end{figure}
In particular, we consider the single user is deployed at $(75\lambda, 100\lambda,0)$, the transmit power is $P_t=5$~dB, and the communication threshold for the estimated user position is $\Gamma_c=5$~dB. The residual error between the estimated user position and the actual user's position is defined as $\Delta_{\theta}$. Fig.~\ref{fig:imperfect_CSI} reveals that when the estimated \ac{csi} error remains below $6^\circ$, the \ac{sinr} of radar detection and user communication is almost unchanged. However, as the estimated error exceeds this threshold and becomes large, the proposed system loses the capability to maintain satisfactory communication performance.

\subsection{Discussion} 

    \subsubsection{Differences between the existing hybrid RIS and our considered HIRS} 

    The existing hybrid \ac{ris} architecture known as STAR-\ac{ris} has been widely considered recently \cite{ni2022star,wuchenyu2021,ahmed2023}. Within the STAR-\ac{ris} framework, each antenna element of the surface splits its incoming signal power into two components. To be specific, one portion of the signal is redirected for reflection within the same region as the incident signal while the rest is transmitted. Therefore, STAR-\ac{ris} exhibits the capability of facilitating both reflection and transmission. In other words, the STAR-\ac{ris} can be viewed as a combination of the reflected-only \ac{ris} and transmitted-only \ac{ris}. 

    However, our considered \ac{hris} differs from the simultaneously transmitting and reflecting (STAR)-\ac{ris}. From the perspective of hardware design, hybrid metasurfaces are designed for operating reflection and reception that allow meta-atoms to both reflect and sense \cite{zhang2022channel,alexandropoulos2021hybrid}. In our \ac{hris}-\ac{dfrc} system, each element of the surface can simultaneously reflect and receive the incoming signals. Specifically, portions of the signal are reflected to the communication user and parts of the impinging signal are received by antennas that are attached to the \ac{rf} chain. We can modify these functionalities by changing the power splitting factor $\boldsymbol{\beta}$ and phase shifts. 

    \subsubsection{Challenges in the HRIS-assisted DFRC system} 
    
There are existing challenges associated with simultaneously transmitting radar and communication signals in an \ac{hris} system.
Firstly, the signals reflected by the \ac{hris} have arrived at both the communication users and the radar detection zone and will interfere with both radar and communication. In this paper, we develop an \ac{ao} algorithm to characterize the trade-off between the radar and communication and propose a sophisticated approach to optimize the power allocation to individual \ac{hris} elements. This optimization strategy seeks to find a balance between radar performance and communication effectiveness in the \ac{hris}-assisted \ac{dfrc} system.

Secondly, based on the inherent characteristics of the \ac{hris}, the \ac{rf} chain is embedded in the \ac{hris}, facilitating the radar echo reception \cite{zhang2022channel,alexandropoulos2021hybrid}. This integration increases the power efficiency of the radar system but decreases the power efficiency of the communication. Considering the hybrid metasurface configuration, the substantial number of elements on the \ac{hris} leads to high computational complexity, which is challenging to optimize and configure. Additionally, there are significant challenges related to the high complexity of the \ac{sdr} when the number of \acp{bs} is large. Some recent research explored finding more memory-efficient and computationally advantageous alternatives to solve the \ac{sdp} problem \cite{Tranter2017,tang2020range,tangbo2020}. In future work, we will consider incorporating \ac{sdr} acceleration to reduce the overall complexity of the \ac{ao} algorithm.

	\vspace{-0.2cm}
	\section{Conclusion}
	\label{sec:Conclusions}
In this work, we proposed an \ac{hris}-assisted \ac{mimo} \ac{dfrc} system, where the \ac{hris} performed reflecting communication signals and receiving radar echo concurrently. With the \ac{sinr} as the evaluation metric of both radar and communication, we characterized the trade-off between radar and communication as a joint optimization problem of the \ac{bs} beamforming design and the \ac{hris} configuration design. Aiming to tackle this problem, we proposed an \ac{ao} approach that consists of the \ac{fgs}-\ac{agd} algorithm for solving the \ac{hris} configuration optimization and an \ac{sdr} technique for the \ac{bs}'s transmitted beam design. Our simulation indicated an apparent trade-off between the performance of the radar and communication while optimizing the joint design of the \ac{bs} and the \ac{hris}. Numerical results demonstrated that the \ac{hris}-assisted system designed by the proposed approach can improve the radar sensing quality and ensure communication compared to the benchmark systems.

The fact that our work has certain underlying assumptions as the \ac{hris}-assisted \ac{dfrc} system is a theoretical concept. Firstly, we derive our model under the assumption of perfect \ac{csi} at the transmitter and the \ac{hris}. However, in practice, this assumption may not hold with the presence of the \ac{csi} estimation error. In future research, the \ac{hris} configuration and \acp{bs} beamforming design considering scenarios where only partial knowledge of the channel is available at the transmitter will be valuable to study. Secondly, in our proposed \ac{hris}-assisted \ac{dfrc} system, we proceed with radar detection in a non-cluttered environment. In more realistic scenarios, the \ac{sinr} of the radar and communication degradation occurs due to clutter. In the future, we intend to investigate radar beamforming adaptations to address cluttered environments, ensuring robust performance in such scenarios.

\ifFullVersion
\vspace{-0.2cm}
\begin{appendix}
	\numberwithin{proposition}{subsection} 
	\numberwithin{lemma}{subsection} 
	\numberwithin{corollary}{subsection} 
	\numberwithin{remark}{subsection} 
	\numberwithin{equation}{subsection}	
	%

	\vspace{-0.2cm}
	\subsection{Proof of Proposition \ref{proposition1}}
	\label{app:Proof_1}
 
First, we separate the reflected matrix $\boldsymbol{\Psi}(\boldsymbol{\beta})$ into power splitting factor $\boldsymbol{\beta}$:~$\mathrm{diag}([\beta_1,...,\beta_N])$. From (\ref{cascaded channel}), the cascaded channel for the $k$-th user is given by 
\begin{equation}
\label{cascaded channel_k}
    \begin{aligned}
    \centering
    \boldsymbol{\hat{h}}_k(\boldsymbol{\beta})=\boldsymbol{h}_k^H \mathrm{diag}([\beta_1,...,\beta_N])\boldsymbol{G}.
    \end{aligned}
\end{equation}

Next, 
by expressing the cascaded communication channel via (\ref{cascaded channel_k}) and combining with (\ref{eq23}), the received power at the user can be recast as (\ref{P_c_sim}).
\begin{figure*}
    \begin{equation}
    \label{P_c_sim}
    \centering \boldsymbol{\hat{h}}_k^H(\boldsymbol{\beta})\boldsymbol{\bar{R}}_k^{(t)}\boldsymbol{\hat{h}}_k(\boldsymbol{\beta})=\boldsymbol{\beta}^T\left(\boldsymbol{h}_k\odot\left(\boldsymbol{G}\boldsymbol{\bar{w}}_{k}^{(t)}\right)\right)\left(\boldsymbol{h}_k\odot\left(\boldsymbol{G}\boldsymbol{\bar{w}}_{k}^{(t)}\right)\right)^H\boldsymbol{\beta},~k=1,...,K.
\end{equation}
\hrulefill
\end{figure*}
Here, we denote $\boldsymbol{c}_3 = \boldsymbol{h}_k\odot\left(\boldsymbol{G}\boldsymbol{\bar{w}}_{k}^{(t)}\right)$ and $\boldsymbol{C}_3 = \boldsymbol{c}_3\boldsymbol{c}_3^H$. Then, the \ac{sinr} at the $k$-th communication user in (\ref{eq23}) is simplified to
\begin{equation}
\label{etac}
    \centering
    \eta_c(\boldsymbol{\beta};k) = \frac{\boldsymbol{\beta}^T\boldsymbol{C}_3\boldsymbol{\beta}}{\boldsymbol{\hat{h}}_k^H\boldsymbol{\bar{R}}^{(t)}\boldsymbol{\hat{h}}_k-\boldsymbol{\beta}^T\boldsymbol{C}_3\boldsymbol{\beta}+\sigma^2}, ~k=1,...,K.
\end{equation}


Compared to (\ref{P_c_sim}), we also transfer the expression of $\eta_r$ into a combination of quadratic forms. 	
We rewrite the cascaded reflected vector $\boldsymbol{\hat{a}}_h$ in the same way as we did in (\ref{cascaded channel_k}), and the decomposed expression of $\boldsymbol{\hat{a}}_h$ is 
\begin{equation}
\label{A_h}
    \centering
    \boldsymbol{\hat{a}}_h^H = \boldsymbol{a}_h^H \mathrm{diag}([\beta_1,...,\beta_N])\boldsymbol{G}.
\end{equation} 

Then, combining with (\ref{A_h}), the quadratic component $\boldsymbol{\hat{a}}_h^H(\boldsymbol{\beta})\boldsymbol{\bar{W}}^{(t)}\boldsymbol{\bar{W}}^{(t)H}\boldsymbol{\hat{a}}_h(\boldsymbol{\beta})$ of the interference power in (\ref{eq18_}) is rewritten by (\ref{A-hRA-h}).
\begin{figure*}[ht]
    \begin{equation}
\label{A-hRA-h}
    \centering
    \boldsymbol{\hat{a}}_h^H(\boldsymbol{\beta})\boldsymbol{\bar{W}}^{(t)}\boldsymbol{\bar{W}}^{(t)H}\boldsymbol{\hat{a}}_h(\boldsymbol{\beta})=\boldsymbol{\beta}^T\left(\boldsymbol{a}_h\odot\left(\boldsymbol{G\bar{W}}^{(t)}\right)\right)\left(\boldsymbol{a}_h\odot\left(\boldsymbol{G\bar{W}}^{(t)}\right)\right)^H\boldsymbol{\beta}.
\end{equation}
\hrulefill
\end{figure*}

Next, we use $\boldsymbol{c}_2$ to replace the $\boldsymbol{a}_h \odot\left(\boldsymbol{G}\boldsymbol{\bar{W}}^{(t)}\right)$, so that the interference power from \ac{hris} to the $(p, q)$ block is written as
\begin{equation}
\label{A_h_c2}
    \centering
    \mathbf{E}\left(|A_r(\boldsymbol{\beta})\boldsymbol{\hat{a}}_h^H(\boldsymbol{\beta})\boldsymbol{x}(n)|^2\right) = |A_r(\boldsymbol{\beta})|^2\boldsymbol{\beta}^T \boldsymbol{C}_2\boldsymbol{\beta},
\end{equation}
where $\boldsymbol{C}_2=\boldsymbol{c}_2{\boldsymbol c}_2^H$ is a Hermitian matrix. Similarly, we separate the \ac{hris}'s received vector $\boldsymbol{\phi}(\boldsymbol{\beta})$ into amplitude vector $1- \boldsymbol{\beta}$. Thus, the cascaded scalar $A_r$ can be rewritten as
\begin{equation}
\label{A-r}
    \centering
    A_r = (1-\boldsymbol{\beta}^T)\boldsymbol{a}_r.
\end{equation}

Now, we can recast the \ac{sinr} of the radar:
\begin{equation}
\label{eta}
    \centering
    \eta_r(\boldsymbol{\beta};p,q) = \frac{(1- \boldsymbol{\beta}^T)\boldsymbol{a}_r\boldsymbol{a}_t^H\boldsymbol{\bar{w}}_{r}^{(t)}\boldsymbol{\bar{w}}_{r}^{(t)H}\boldsymbol{a}_t\boldsymbol{a}_r^H(1- \boldsymbol{\beta})}{(1-\boldsymbol{\beta}^T)\boldsymbol{a}_r\boldsymbol{\beta}^T \boldsymbol{C}_2\boldsymbol{\beta}\boldsymbol{a}_r^H(1- \boldsymbol{\beta})+\sigma^2}.
\end{equation}

By substituting $\boldsymbol{C}_1$ for the $\left({\boldsymbol a}_t^H\boldsymbol{\bar{w}}_{r}^{(t)}\boldsymbol{\bar{w}}_{r}^{(t)H}\boldsymbol{a}_t\right)\boldsymbol{a}_r\boldsymbol{a}_r^H$,the $\eta_r$ is consequently reduced to:
\begin{equation}
\label{etar}
    \centering
    \eta_r(\boldsymbol{\beta};p,q) = \frac{(1- \boldsymbol{\beta}^T)\boldsymbol{C}_1(1- \boldsymbol{\beta})}{(1- \boldsymbol{\beta}^T)\boldsymbol{a}_r\boldsymbol{\beta}^T \boldsymbol{C}_2\boldsymbol{\beta}\boldsymbol{a}_r^H(1- \boldsymbol{\beta})+\sigma^2}.
\end{equation}
This concludes the proof.

\end{appendix}	
\fi 
 
	\bibliographystyle{IEEEtran}
	\bibliography{IEEEabrv,mybib}
\vspace{-1.2cm}
\begin{IEEEbiography}
[{\includegraphics[width=1in,height=1.25in,clip,keepaspectratio]{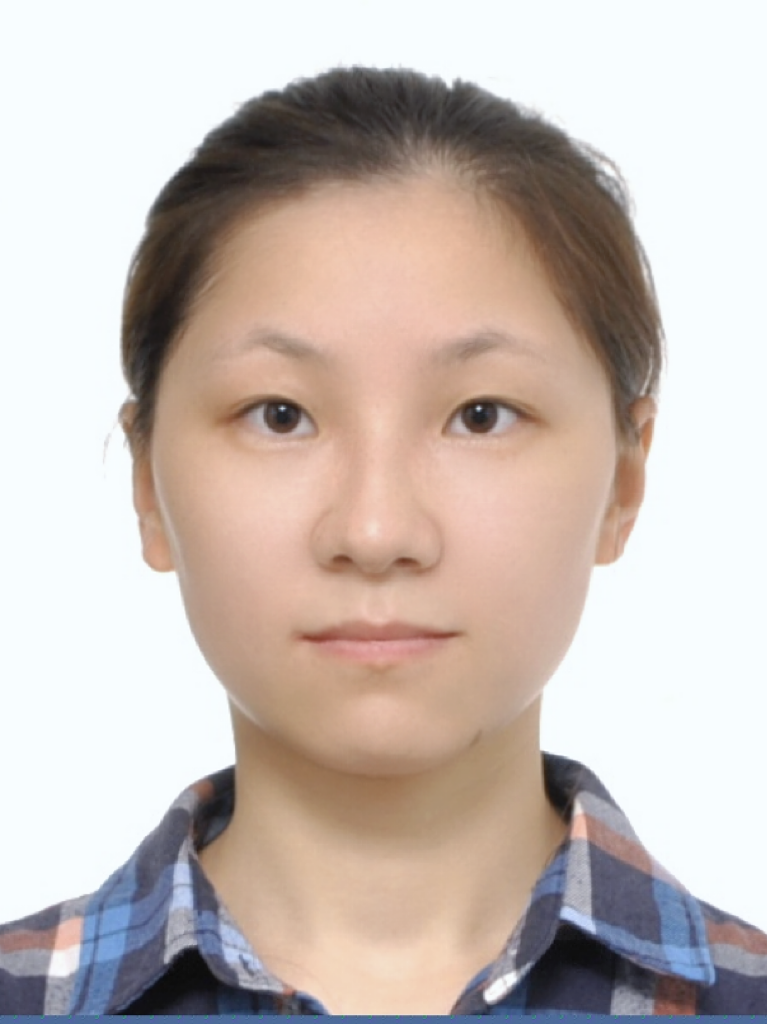}}]{Zhuoyang Liu}
    (Student Member, IEEE) received the B.E. degree from Wuhan University, Wuhan, Hubei, China, in 2020. She is currently pursuing the Ph.D. degree in electromagnetic science with the Key Laboratory of Information Science of Electromagnetic Waves, Fudan University, Shanghai, China. Her research interests include signal processing, reconfigurable intelligence surfaces, and the combination of artificial intelligence and electromagnetic waves.
\end{IEEEbiography}
\vspace{-1.2cm}	
\begin{IEEEbiography} 
[{\includegraphics[width=1in,height=1.25in,clip,keepaspectratio]{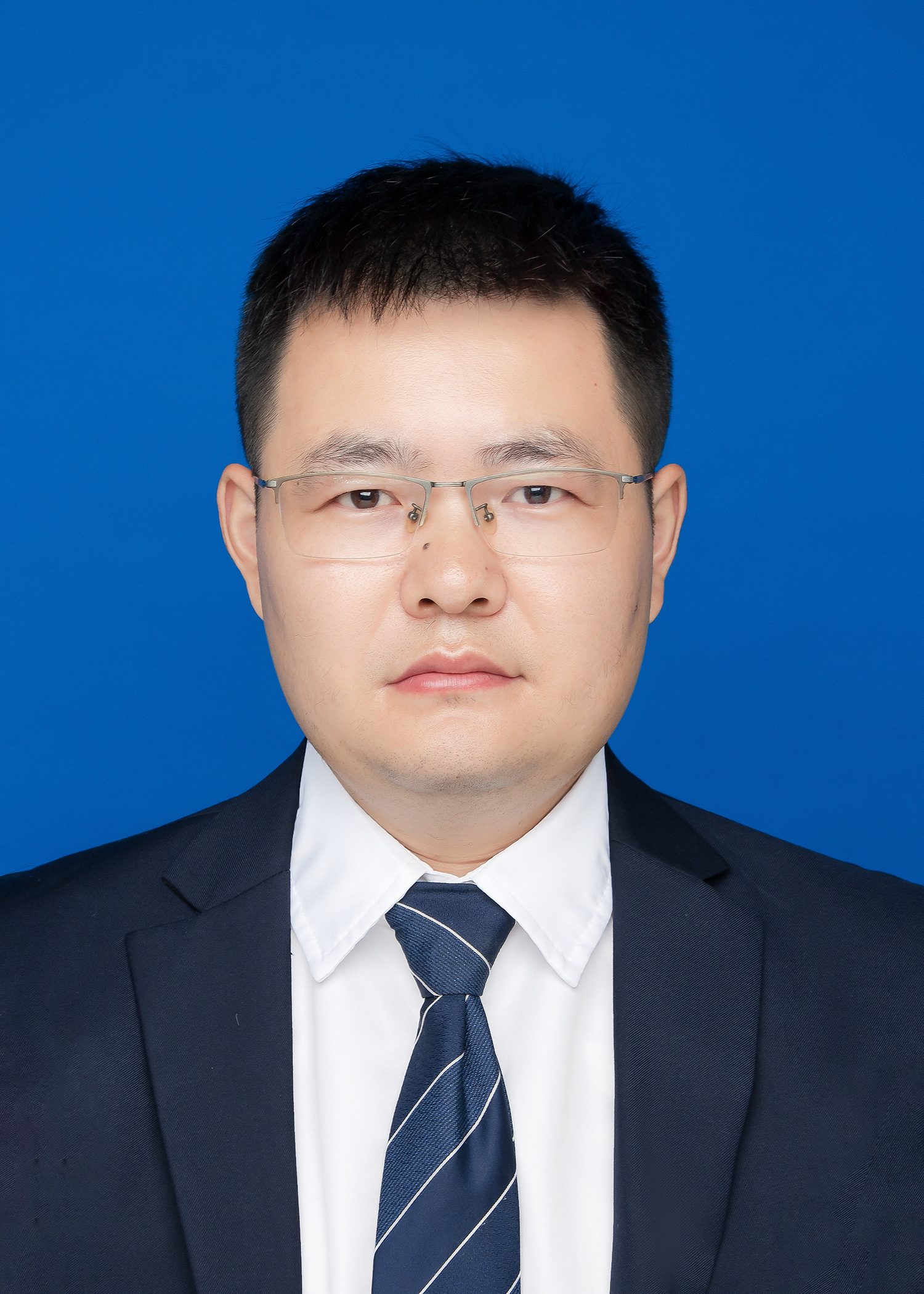}}]{Haiyang Zhang}
	(Member, IEEE) received the B.S. degree in communication engineering from Lanzhou Jiaotong University, Lanzhou, China, in 2009, the M.S. degree in information and communication engineering from the Nanjing University of Posts and Telecommunications, Nanjing, China, in 2012, and the Ph.D. degree in information and communication engineering from Southeast University, Nanjing, China, in 2017. He is currently a Professor with the School of Communications and Information Engineering, Nanjing University of Posts and Telecommunications. His research interests include 6G near-field MIMO communications, deep learning and sampling theory.  
\end{IEEEbiography}	
\vspace{-1.2cm}
\begin{IEEEbiography}[{\includegraphics[width=1in,height=1.25in,clip,keepaspectratio]{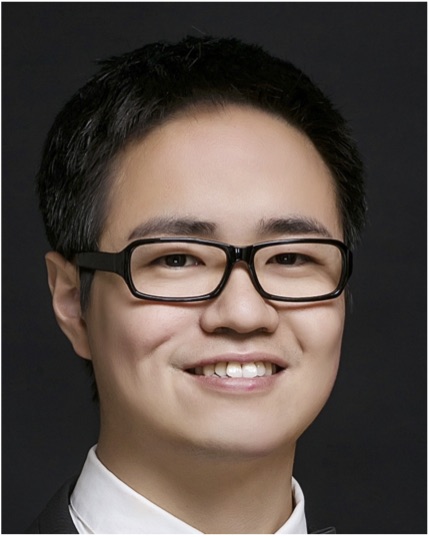}}]
	{Tianyao Huang} received the B.S. degree in 2009 in telecommunication engineering from the Harbin Institute of Technology, Heilongjiang, China, and the Ph.D degree in 2014 in electronics engineering from the Tsinghua University, Beijing, China, respectively. 
	He is currently a Professor with the School of Computer and Communication Engineering, University of Science and Technology Beijing, Beijing, P.R. China. From 2017 to 2023, he was with the Intelligence Sensing Lab, Department of Electronic Engineering, Tsinghua University, as an assistant professor.
	His current research interests include signal processing, compressed sensing, and joint radar communications system design.
\end{IEEEbiography} 
\vspace{-1.2cm}
\begin{IEEEbiography} 
[{\includegraphics[width=1in,height=1.25in,clip,keepaspectratio]{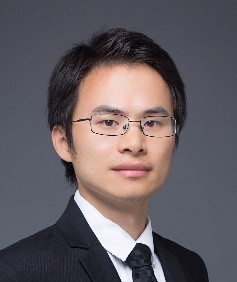}}]{Feng Xu} 
(S’06-M’08-SM’14) received the B.E. with honor in Information Engineering from Southeast University, Nanjing, China and the Ph.D. with honor in Electronic Engineering from Fudan University, Shanghai, China, in 2003 and 2008, respectively. He currently serves as vice dean of the School of Information Science and Technology and director of the MoE Key Lab for Information Science of Electromagnetic Waves. 
His current research interests include electromagnetic scattering theory, SAR information retrieval, and advanced radar systems. 
\end{IEEEbiography}	
\vspace{-1.2cm}
\begin{IEEEbiography}[{\includegraphics[width=1in,height=1.25in,clip,keepaspectratio]{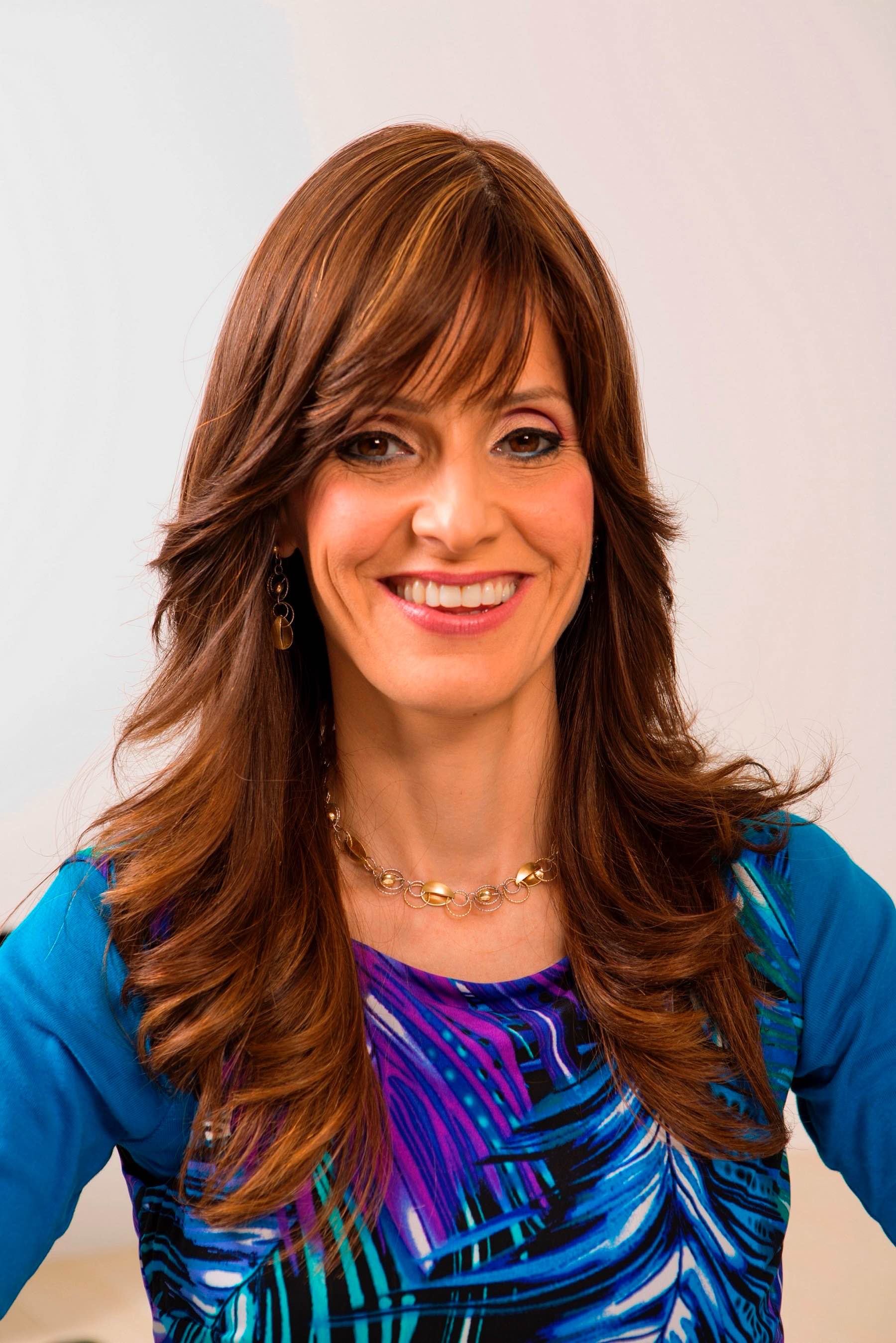}}]
	{Yonina C. Eldar} (S’98–M’02–SM’07–F’12) received the B.Sc. degree in physics and the B.Sc. degree in electrical engineering from Tel-Aviv University, Tel-Aviv, Israel,	1995 and 1996, respectively, and the Ph.D. degree in electrical engineering and computer science from the Massachusetts Institute of Technology (MIT), Cambridge, MA, USA, in 2002.	
	She is currently a Professor with the Department of Mathematics and Computer Science, Weizmann Institute of Science, Rehovot, Israel. Her research interests include statistical signal processing, sampling theory and compressed sensing, learning and optimization methods, and their applications to biology, medical imaging and optics.
\end{IEEEbiography}

\end{document}